\documentclass{aa}

\usepackage{graphicx}
\usepackage{txfonts}
\usepackage{url}
\usepackage{hyperref}
\hypersetup{
     breaklinks,
     colorlinks   = true,
     citecolor    = blue,
     urlcolor     = blue,
     linkcolor    = blue
}

\usepackage{placeins}
\usepackage[rightcaption]{sidecap}
\usepackage{makecell}
\newcommand{\rot}[1]{\rotatebox{90}{#1}}

\begin{document} 

   \title{Impact of selection criteria on the structural parameters \\of the Galactic thin and thick discs}
   \titlerunning{Impact of selection criteria on the structural parameters of the Galactic discs}

   \author{S. Alinder\inst{1},
          T. Bensby \inst{1}
          \and
          P. J. McMillan\inst{2}
          }

   \institute{Lund Observatory, Division of Astrophysics, Department of Physics, 
            Lund University, Box 118, SE-221\,00 Lund, Sweden \and
         School of Physics \& Astronomy, University of Leicester, University Road, Leicester LE1 7RH, UK.\\
              \email{simon.alinder@fysik.lu.se}
             }

   \date{Received 13 November 2025; accepted 27 May 2026}
 
  \abstract
   {
    The dual nature of the Milky Way disc is well established. It contains a thick and a thin disc that differ in chemical, kinematic, structural, and spatial properties.
    Simultaneously, there is significant overlap in the distributions of these properties, especially at higher metallicities.
    Accurately distinguishing between these major structural components is crucial for understanding the formation and evolution of the Milky Way.
   }
   {
   We aim to investigate how the various classification methods for categorising stars into the thick and thin disc populations influence the determination of the structural properties of the two discs.
   }
   {
    We applied five different selection methods to define samples of stars that are likely members of either the thick or the thin disc.
    Two methods use cuts in the [$\alpha$/Fe]-[Fe/H] and [Mg/Mn]-[Al/Fe] planes; one method uses a dynamical separation in $J_\phi $-$ J_Z$ space; one uses an age-based cut; and the last method uses a kinematic likelihood method.
    For each method, we derived the relative density profiles of each disc component as functions of height above the Galactic plane and Galactocentric radius, and fitted these to a simple two-exponential disc model. For our analysis, we used red giant stars from APOGEE DR17 and stellar ages from astroNN.
   }
   {
    Methods based on abundance or age data produce the cleanest separations, while kinematic and dynamical methods suffer higher contamination due to difficulties in separating well-mixed populations.
    The thin disc scale heights show a clear flaring as they increase with radius, while the thick disc stays approximately constant at around 1\,kpc over most radii for all methods.
    All methods find the thin disc to have a longer scale length than the thick disc, with the difference being greatest for the chemical selection methods.
    A thick disc scale length of 2.0\,kpc yields a thin disc scale length of between 2.3 and 3.0\,kpc.
   }
   {
    The method chosen to distinguish between Galactic components can significantly impact the view of the Galaxy.
    Abundance-based approaches offer a clear separation but require spectroscopy, which is not as widely available, while kinematic and dynamical methods work on larger samples but suffer from less clear separations.
    Every method recovers a longer scale length for the thin disc than the thick disc, unlike the geometric method, used with luminosity profiles in external galaxies.
   }


   \maketitle


\section{Introduction}
The two-component model of the Milky Way disc, comprising a thick and a thin disc, has been a cornerstone of Galactic structure studies for decades \citep[see e.g.][]{juric_milky_2008, chang_information_2011, hayden_ambre_2017, kawata_milky_2026}. Such a model typically relies on fitting two exponential profiles to observed stellar density distributions; \cite{gilmore_new_1983} observed stars in the direction of the Galactic south pole and found they needed two exponentials to fit their data, one with a scale height of 300\,pc and one with a scale height of 1350\,pc. They are now known as the thin and thick discs, respectively.

The thick and thin discs have distinct structural, kinematic, and chemical characteristics. The thick disc reaches higher above the Galactic plane and is less radially extended \citep{bensby_first_2011, bovy_spatial_2012}. It contains stars that are older, have lower metallicities, and are richer in $\alpha$-elements compared to the thin disc stars \citep{fuhrmann_nearby_1998, bensby_elemental_2003, reddy_chemical_2003, katz_radial_2021}. Kinematically selected samples of thick disc stars have a higher velocity dispersion and a larger asymmetric drift than thin disc stars \citep[e.g.][]{casetti-dinescu_three-dimensional_2011}, which has been used to preferentially select stars from either disc \citep{fuhrmann_nearby_1998, gilmore_new_1983, bensby_elemental_2003, binney_galactic_2008}. 

Analogous dual disc structures have been identified in external galaxies as breaks in their luminosity profiles when seen edge-on \citep[e.g.][]{comeron_breaks_2012}. In contrast to the Milky Way, the thick discs in these external galaxies always have scale lengths longer than the thin disc ones. \cite{minchev_relationship_2017} claim that the observed thick discs are many smaller mono-age populations that are stacked and flared. They use chemodynamical modelling to argue that discs that form from the inside out produce nested flared populations with the same age that, when summed, mimic the geometry of a thick disc without requiring a separate disc component. In this view, there should be a gradient in age in the geometrically defined thick disc as the older populations have flared more, leaving a sequence of ages in the radial direction \citep[see also][]{martig_radial_2016}. The data available for external galaxies are more limited, so fitting one or two luminosity profiles to the observed height above the mid-plane is often the only viable means of distinguishing disc components beyond the Milky Way.

How the Milky Way came to possess two discs has been discussed for many years (see, e.g. the review by \citealt{rix_milky_2013}). The two major theories for the formation mechanism of the thick disc are in situ formation and ex situ formation. In an in situ scenario, high turbulence naturally leads to large scale-heights \citep{bournaud_thick_2009} or radial migration causes stars to reach greater $Z$-heights \citep{loebman_genesis_2011}. In an ex situ scenario, dynamical heating or gas-rich mergers, such as the Gaia-Enceladus-Sausage event, heat a pre-existing thin disc or proto-disc and trigger centrally concentrated starbursts \citep{helmi_merger_2018, belokurov_co-formation_2018}. The current consensus suggests both processes played roles: an early, rapid in situ buildup supplemented by later merger-driven heating \citep[e.g.][]{pinna_stellar_2024}. Following the formation of the thick disc, the Milky Way entered a more quiescent phase characterised by a slower, inside-out growth of the thin disc \citep{haywood_milky_2016}. Semi-analytic and chemical evolution models, especially two-infall scenarios \citep{chiappini_chemical_1997}, propose a gap in the star formation of $\sim\,1$\,Gyr, followed by a phase of gradual accretion of enriched gas and slower star formation that builds an extended, kinematically colder thin disc \citep{haywood_milky_2016, katz_radial_2021}. There are chemical evolution models that do not include this gap and are still able to produce thin and thick discs \citep[e.g.][]{schonrich_origin_2009, schonrich_understanding_2017}. In either scenario, accreted gas is mixed with the enriched gas already present, which, together with dynamical processes like the radial migration of stars, results in the metallicity gradients and age-radius correlations that are observed today \citep{cerqui_chemical_2025}.

We currently have access to a wealth of high-quality data of stars in the Milky Way thanks to the European Space Agency's astrometric \textit{Gaia} mission \citep{gaia_collaboration_gaia_2016, gaia_collaboration_gaia_2016-1, gaia_collaboration_gaia_2018, gaia_collaboration_gaia_2021, gaia_collaboration_gaia_2023} and spectroscopic surveys like APOGEE \citep{majewski_apache_2017, abdurrouf_seventeenth_2022}, the \textit{Gaia}-ESO survey \citep{gilmore_gaia-eso_2022, randich_gaia-eso_2022}, GALAH \citep{buder_galah_2025}, and LAMOST \citep{cui_large_2012}. With these surveys, the Milky Way disc can be analysed in great depth. However, even with these data, categorising any star as belonging to the thick or thin disc is far from straightforward. It is possible to categorise stars based on geometric, kinematic, dynamical, chemical, or chronological criteria, each using different aspects of the stellar populations \citep{martig_radial_2016}. The choice of selection depends on the scientific question being considered, but the selections are not always equivalent and can lead to differing interpretations. For example, the Galaxy has stars at high $Z$ altitudes in the outer part of the disc. Under a geometric interpretation, these are thick disc stars, and the Galaxy therefore has an extended thick disc. Under a chemical interpretation, however, those stars are classified as thin disc stars, and the Galaxy has a short thick disc and a flared thin disc \citep{bensby_first_2011}.

It is important to investigate whether different definitions of these components lead to different selections of groups of stars with different properties. If different methods yield inconsistent classifications, this would raise questions about how we interpret the structure of the disc and lead to possible confusion when structures defined in different ways are called the same thing. In this paper, we systematically compare two chemical, one kinematic, one dynamical, and an age-based selection method by using them to fit a simple model of the vertical density profile of the Milky Way and the density fraction of each disc component in the plane of the disc. We also compute the surface density fractions of the two discs and their scale lengths. The different methods for categorising stars we examine are:
\begin{itemize}
    \item Abundance-based selections, which look at stars in chemical spaces and try to find distinct populations. Several different elements can and have been used; \cite{buckley_chemical_2024} explore which elemental abundances are most effective for distinguishing components of the disc from each other. $\alpha$ elements like magnesium are frequently used, as there are two populations discernible in an $\alpha$-metallicity plot \citep{fuhrmann_nearby_1998, bensby_elemental_2003, reddy_chemical_2003}.
    
    \item A kinematic selection, which relies on stellar velocities in three dimensions. Thick disc stars have higher velocity dispersions and distinct orbital characteristics compared to thin disc stars \citep[e.g.][]{bensby_elemental_2003}.
    
    \item A dynamical selection, which goes further than the kinematic method by incorporating both positions and velocities, evaluated within the context of a model of the Galactic gravitational potential. This allows for the computation of quantities such as orbital actions and energy, offering a more physically grounded classification framework \citep{robin_self-consistent_2022}.
    
    \item An age-based selection, which is based on stellar ages and relies on the assumption that thick disc stars formed earlier in the history of the Galaxy \citep{haywood_age_2013, bensby_exploring_2014}. Stellar ages are notoriously difficult to determine with high precision for large samples, making this method challenging to apply in practice, at least until accurate age estimates become available for a larger number of stars.
\end{itemize}

We test methods that are used frequently, that are used infrequently, and that could potentially be used in the future, with the goal of examining how each method influences the inferred properties of the thick and thin disc populations, thereby clarifying to what extent methodological choices affect scientific conclusions about the structure of the Galaxy.
In Sect.~\ref{section:data}, we present the data and how the stellar sample was selected, and in Sect.~\ref{section:selections}, we describe the methods we used to categorise the thin and thick disc samples. The model we use to compute the scale parameters of the discs is presented in Sect.~\ref{section:methods}, followed by the results in Sect.~\ref{section:results}. Lastly, the results are discussed in Sect.~\ref{section:discussion}, and in Sect.~\ref{section:conclusions} we provide a summary and conclusion.

\section{Data}\label{section:data}

\begin{figure*} 
\centering
\resizebox{0.9\hsize}{!}{
\includegraphics{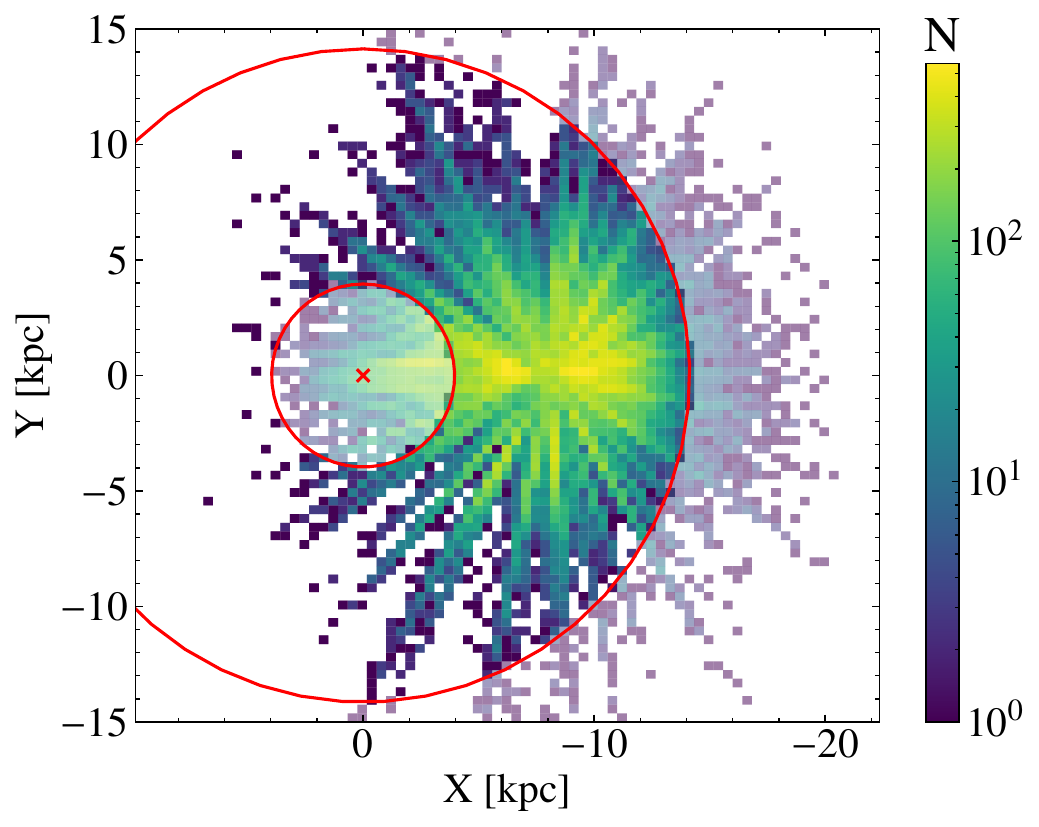}
\includegraphics{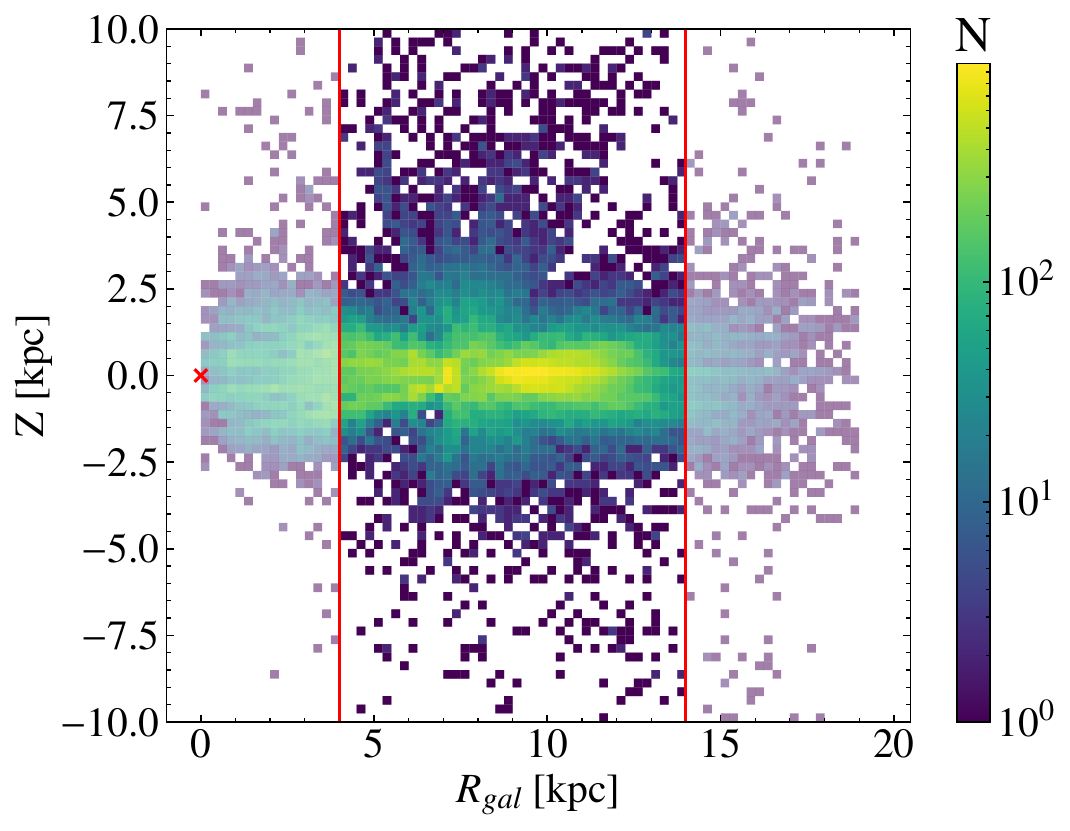}}
  \caption{Distribution of the selected stars for this study in the X-Y plane (\textit{left}) and the $ R_{\mathrm{gal}} $-Z plane (\textit{right}).
  Regions outside our radial limits, shown with red lines at $R_{\mathrm{gal}} = 4$ kpc and $R_{\mathrm{gal}} = 14$ kpc, are semi-transparent.
  The Galactic centre is marked with a red cross.
          }
     \label{fig:location}
\end{figure*}

We used the Main Red Star Sample data from APOGEE DR17 \citep{abdurrouf_seventeenth_2022}, which was constructed with the intention of having a selection function that is easy to understand.
It only contains red giants, which means that it is a high-luminosity population that probes a large part of the Galaxy, allowing us to probe the thin and thick discs well beyond the solar neighbourhood.
Furthermore, the APOGEE dataset is large, with many different elemental abundances, and as such, it is frequently used by the community, which makes our results easier to compare against those of others.
We applied several quality cuts to ensure that our sample of red giant stars is of high quality (see Appendix \ref{appendix:data} for details on how we processed the APOGEE data). It should be noted that the inconsistencies in some abundances in the APOGEE data, notably  [Al/Fe], which is important for one of our selection methods, are only present for dwarf stars and not the red giants \citep{jonsson_apogee_2020}.

To evaluate the proposed selection methods, we fitted a simple axisymmetric model of the Galactic disc to our data. We restricted our data to only include stars with Galactocentric radial distances ($ R_{\mathrm{gal}} $) between 4\,kpc and 14\,kpc, to ensure the model is valid. The inner edge is restricted by the presence of the Galactic bar, which is not an axisymmetric feature \citep{wegg_structure_2015}, and the outer edge is restricted by the limit where our model breaks down due to low star counts.
Top-down X-Y and $ R_{\mathrm{gal}} $-Z visualisations of the distribution of our data are shown in Fig.~\ref{fig:location}.

Even though the selected sample is large and contains stars from across the Galaxy, it is not necessarily representative of the true population of the Galaxy. We needed to consider the selection function to obtain a representative sample. The selection function expresses the probability that a star is included in a survey given the survey's magnitude limits and the location of the star. To find the selection function for our sample, we built on the work done in the \texttt{gaiaunlimited} package\footnote{\url{https://github.com/gaia-unlimited/gaiaunlimited}} \citep{cantat-gaudin_uniting_2024} by using a modified version of the \texttt{apogee\_sf} method that is provided for this purpose.
Our modifications allowed us to compute a selection function for 85\,\% of our sample instead of 63\,\% as in the original implementation.
See Appendix \ref{appendix:selection} for details on this process.

Some of the selection methods we used require the [Mg/Fe], [Mn/Fe], and [Al/Fe] abundance ratios. The sample was therefore limited to stars that have these values. This left a `basic' sample of $106\,163$ stars, which was used for all methods except those which require additional data not included in APOGEE DR17.
For instance, the dynamical method requires information about the orbits of the stars.
For this method, we computed the actions for each star in our sample using \texttt{AGAMA} (Action-based Galaxy Modelling Architecture, \citealt{vasiliev_agama_2019}) with the \texttt{best} potential from \cite{mcmillan_mass_2017}\footnote{
        The galactocentric coordinate system used in this study places the Sun on the negative $X$-axis ($\phi = 180^\circ$) at a distance of 8.122 kpc and a height of 20.8 pc, with the $Y$-axis in the same direction as $l=90^{\circ}$, and the $Z$-axis in the same direction as $b=90^{\circ}$. Galactic azimuth, $\phi$, decreases in the direction of Galactic rotation, and the Sun has velocity components $V_{R, \sun}=-12.9, V_{\phi, \sun}=-245.6, V_{Z, \sun}=-7.78$ km\,s$^{-1}$.
        \citep{reid_proper_2004, drimmel_solar_2018, gravity_collaboration_detection_2018, bennett_vertical_2019}. For the computations and definitions of coordinates, we used Astropy v7.0 \citep{astropy_collaboration_astropy_2022}.
}.
These data were added as additional columns onto the `basic' sample, meaning the number of stars in the sample was unchanged.
The age-based method requires stellar ages, which we got estimates of from \cite{mackereth_dynamical_2019} via the astroNN catalogue for APOGEE DR17. This is a catalogue containing the results from running the \texttt{astroNN} deep-learning code \citep{leung_deep_2019} on APOGEE spectra to estimate stellar ages and other properties like elemental abundances and distances. We selected stars with a relative age uncertainty not exceeding 50\%, which limits the sample size for this method to 90\,387 stars.
This sample was a strict subsample of the `basic' sample above.

\section{Selection methods} \label{section:selections}

Each method we used categorises stars by labelling them as thin disc, thick disc, or halo. The halo category is intended to contain all non-disc stars in the sample. This often means this category includes halo stars, accreted stars, or peculiar disc stars with properties that separate them from other disc stars. Table~\ref{tab:cat numbers} contains a summary of the total number of stars categorised by each method as thin disc, thick disc, or halo.

\begin{table}
\caption{Number of stars selected for each component for each method.}
\small
\centering
\setlength{\tabcolsep}{3mm}
\begin{tabular}{cccc}
\hline \hline
Selection method & Thin disc & Thick disc & Halo \\
\hline
\noalign{\smallskip}
[Mg/Mn] - [Al/Fe] & $8.4 \times 10^4$ & $1.9 \times 10^4$ & $1.9 \times 10^3$ \\
{[Mg/Fe]} - [Fe/H] & $7.7 \times 10^4$ & $2.7 \times 10^4$ & $1.9 \times 10^3$ \\
Dynamical & $7.9 \times 10^4$ & $2.6 \times 10^4$ & $1.0 \times 10^3$ \\
Age & $7.3 \times 10^4$ & $1.7 \times 10^4$ & $ 1.1 \times 10^3 $ \\
Kinematic & $1.4 \times 10^4$ & $5.7 \times 10^3$ & $1.0 \times 10^3$ \\
\hline
\end{tabular}
\label{tab:cat numbers}
\end{table}

We compared the selections by putting the stars selected as thin or thick discs by each method in a figure with an [Mg/Mn]-[Al/Fe] plane and a [Mg/Fe]-[Fe/H] plane.
These comparisons are made using chemical data because those properties are stable over time, unlike the positional or kinematic properties, and directly address the origin of the stars.
Using age data could provide a clear way to distinguish the disc components as suggested by results from, for example, \cite{bensby_exploring_2014}, \cite{martig_radial_2016}, and \cite{xiang_time-resolved_2022}, but would be less useful here because the age data from astroNN have larger uncertainties than the abundance data, which blurs the distinction.
Further results of the different methods than those presented below, including the distributions in the X-Y plane, $ R_{\mathrm{gal}} $-Z, [Mg/Fe]-[Fe/H], [Mg/Mn]-[Al/Fe], $J_\phi$-$J_z$, Age-[Mg/Fe], and Toomre plot planes, are shown in Figs.~\ref{fig:mgmnalfe results full}-\ref{fig:age results full}.

\subsection{Selection using the [Mg/Mn]-[Al/Fe] plane} \label{section:chem_haywood}

\begin{figure*}
\centering
\includegraphics[width=0.45\hsize]{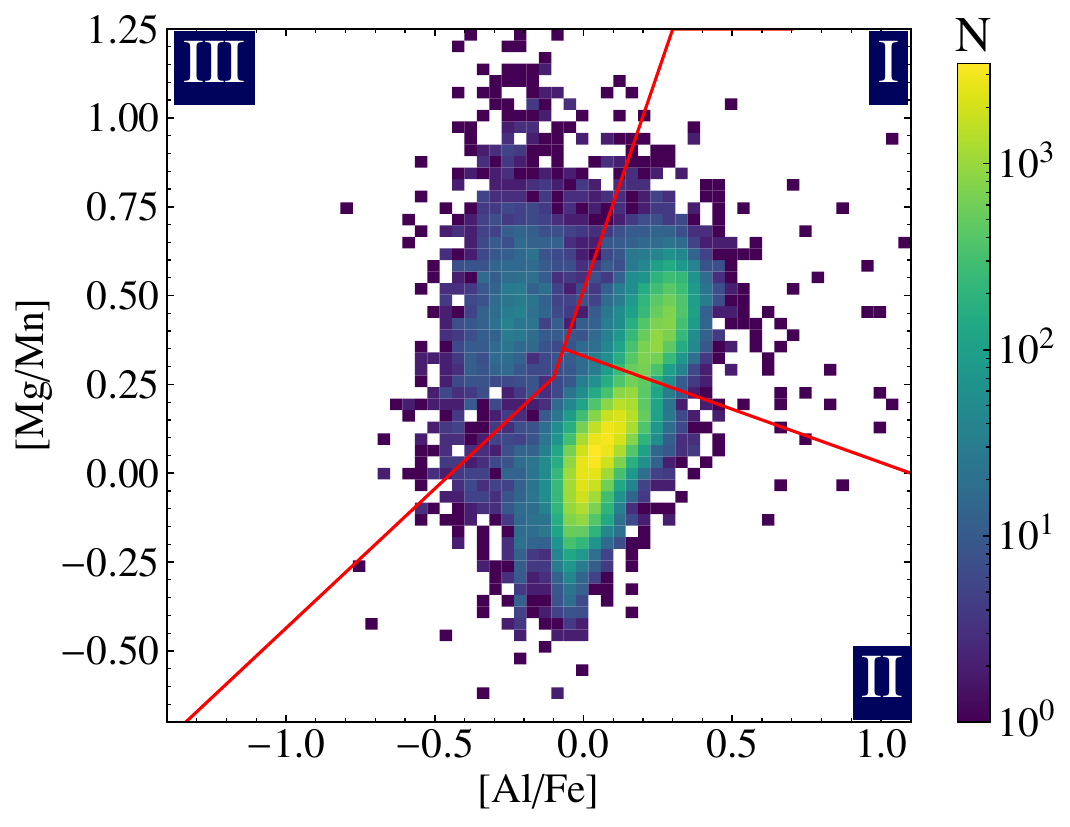}
\includegraphics[width=0.45\hsize]{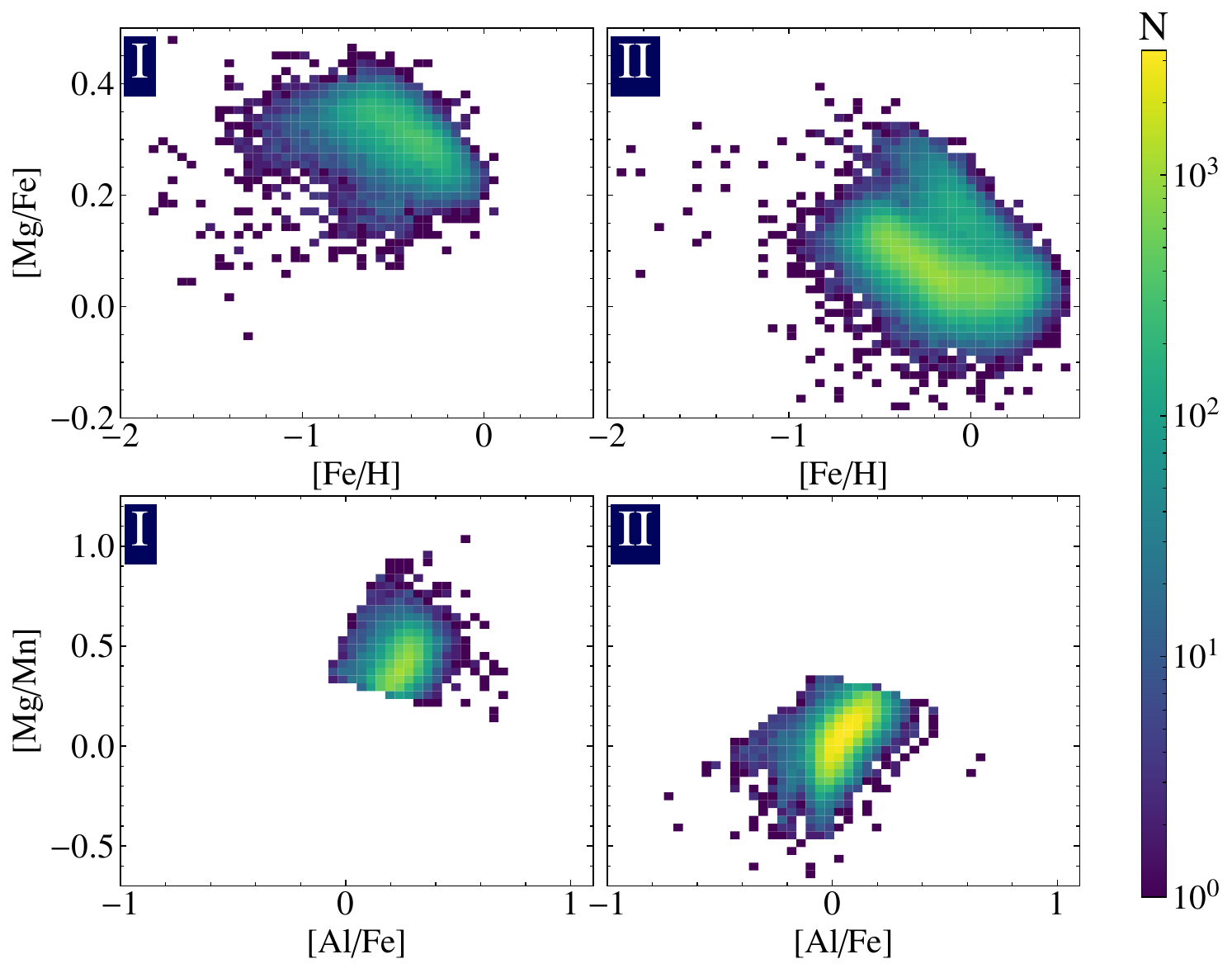}
  \caption{How the accreted population was split from the thin and thick discs using the [Mg/Mn]-[Al/Fe] plane.
  \textit{Left}: Number density in the [Mg/Mn]-[Al/Fe] plane, divided into different regions marked with red lines. Region I is the thick disc, region II is the thin disc, and region III is the halo or accreted stars.
  \textit{Right}: Distribution of stars from the selection:  [Mg/Fe]-[Fe/H] (top row) and [Mg/Mn]-[Al/Fe] (bottom row) for the thick disc (left column) and the thin disc (right column).
          }
     \label{fig:mgmnalfe}
\end{figure*}

One of the most promising methods for separating stars into Galactic components is based on the distribution in the [Mg/Mn]-[Al/Fe] plane.
This relies on the fact that the interstellar medium is enriched with magnesium primarily via type II core-collapse supernovae (SNe\,II), while manganese is an iron-peak element primarily produced by thermonuclear type Ia supernovae (SNe\,Ia). Because SNe\,II are caused by massive stars with short lifetimes, manganese abundance will lag magnesium abundance, allowing [Mg/Mn] to act as an indicator of time of formation \citep{nomoto_nucleosynthesis_2013}. Aluminium enhances the effect by being mostly produced in massive stars, with yields that rise with metallicity and therefore correlate with intense, sustained star formation \citep[and references therein]{nordlander_non-lte_2017}.
Environments with slow, prolonged chemical evolution, such as dwarf galaxies, remain at a lower metallicity for longer and therefore have systematically lower [Al/Fe] \citep{nissen_two_2010, hawkins_using_2015, das_ages_2020}.

This method builds on the work of \cite{hawkins_using_2015} and \cite{das_ages_2020}, who identify the separation between accreted stars and disc stars in this plane.
\cite{das_ages_2020} find the low-[Al/Fe] part to be mostly pure accreted stars in the high-[Mg/Mn] area and Al-poor thin disc stars in the low-[Mg/Mn] area. \cite{horta_evidence_2021} show this low-[Al/Fe] and high-[Mg/Mn] area to be a mix of accreted and unevolved in situ stars. Because this method cannot differentiate between accreted and unevolved in situ stars, we treated the two as the same group. We selected three regions in this plane to use as the thick disc, thin disc, and halo. The left panel in Fig.~\ref{fig:mgmnalfe} shows the lines we draw in the [Mg/Mn]-[Al/Fe] plane to separate the accreted stars, leaving us with the disc stars in the high-[Al/Fe] region. \cite{das_ages_2020} and \cite{feuillet_old_2022} find a population of old low-Al thin disc stars in this plane, so the line separating the in situ stars and the accreted stars is bent to include them. The disc population shows two over-densities, which \cite{hawkins_using_2015} identified with the thick disc (Region I), centred around ([Mg/Mn] $ \approx 0.5$, [Al/Fe] $ \approx 0.3$), and the thin disc (Region II), centred around ([Mg/Mn] $ \approx 0.0$, [Al/Fe] $ \approx 0.0$).

The results of this selection are shown in the smaller panels in Fig. \ref{fig:mgmnalfe}, with the [Mg/Fe]-[Fe/H] and [Mg/Mn]-[Al/Fe] planes for the thick and thin disc selections in the left and right columns, respectively. The thick disc, Region I, forms a short continuous sequence in [Mg/Fe]-[Fe/H], and reaches [Fe/H] $\approx 0$. There is also a small group of stars below the sequence at ([Mg/Fe] $\approx 0.18$, [Fe/H] $\approx -0.6$), which would traditionally be classified as belonging to the low-$\alpha$ thin disc. The thin disc, Region II, has two sequences, one at lower [Mg/Fe] and a smaller, less populated one at higher [Mg/Fe], which would traditionally be classified as belonging to the high-$\alpha$ thick disc.

\subsection{Selection using the [$\alpha $/Fe]-[Fe/H] plane} \label{section:chem_simple}

\begin{figure*}
\centering
\includegraphics[width=0.45\hsize]{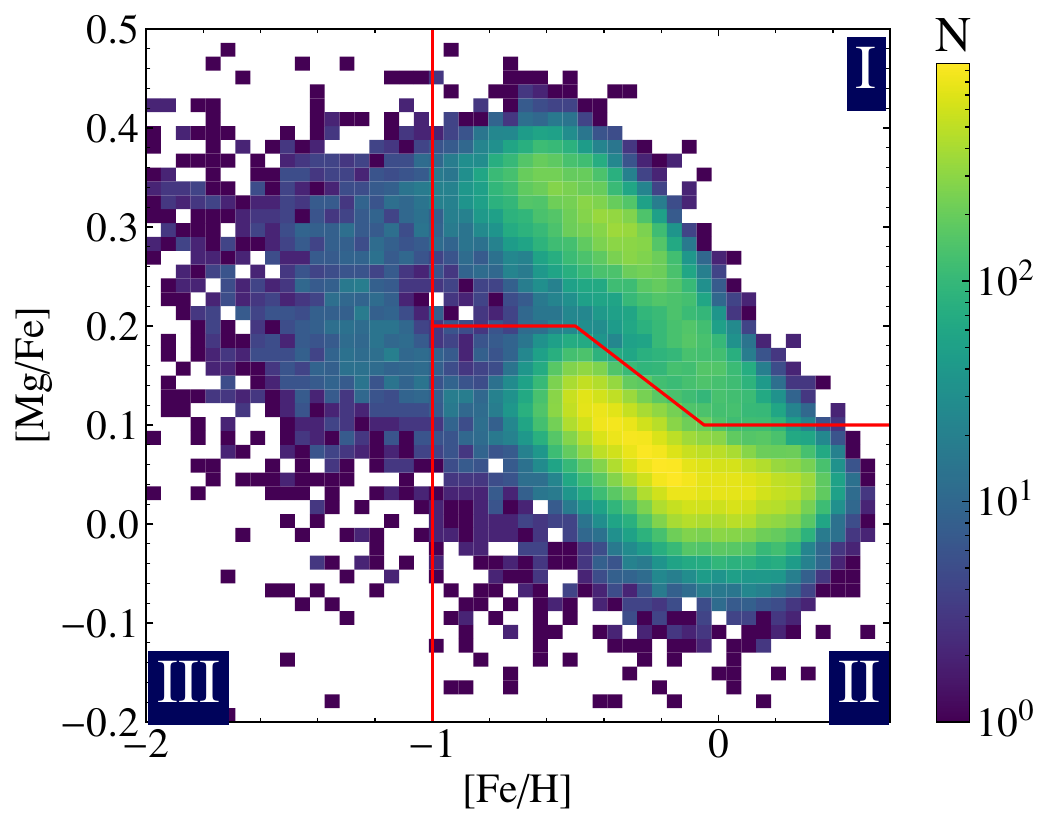}
\includegraphics[width=0.45\hsize]{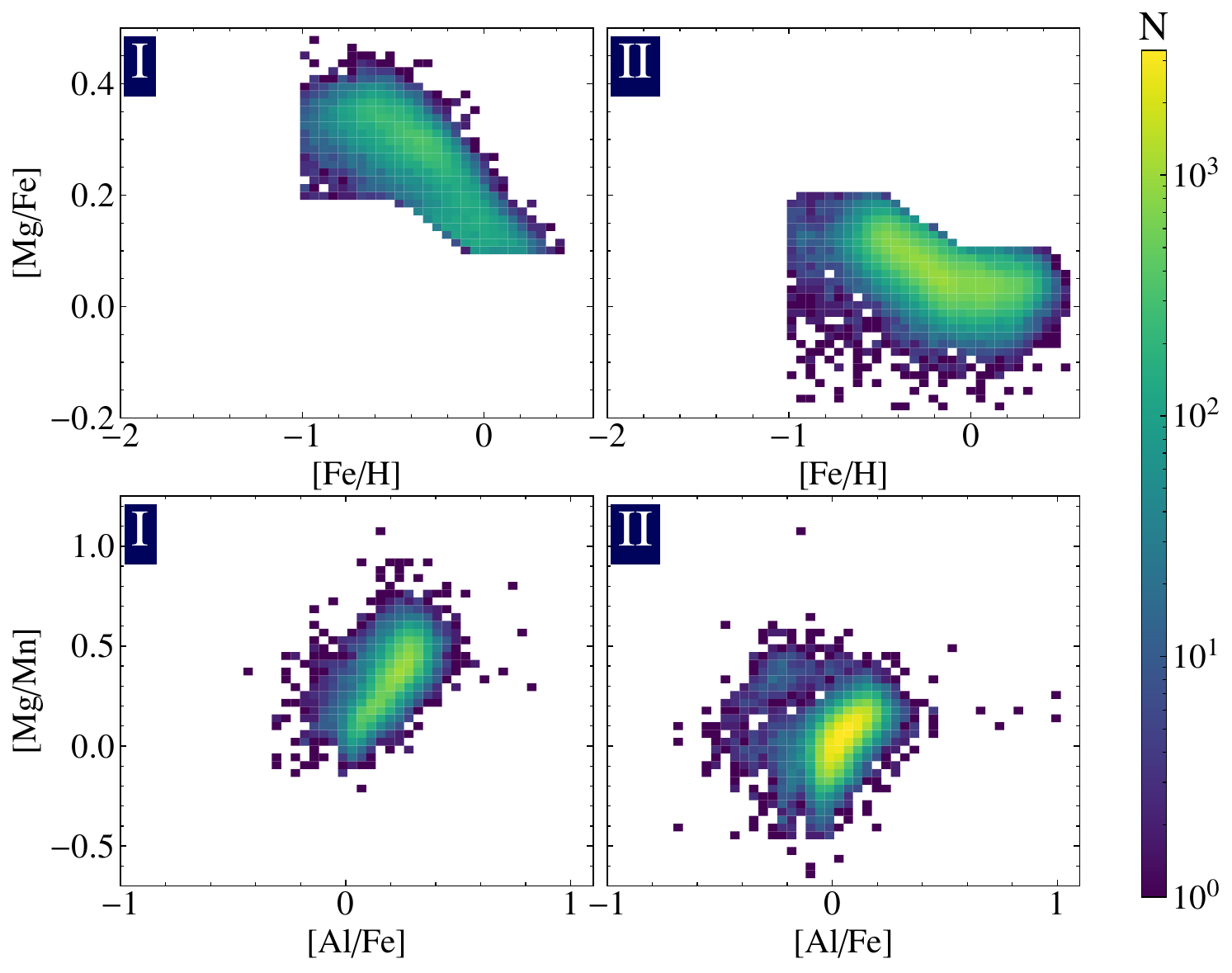}
  \caption{How the [Mg/Fe]-[Fe/H] plane was split into the thin disc, thick disc, and halo.
  \textit{Left}: Number density in the [Mg/Fe]-[Fe/H] plane. Region I is the thick disc, Region II the thin disc, and Region III accreted and halo stars, separated by red lines.
  \textit{Right}: Distribution of stars from the selection: [Mg/Fe]-[Fe/H] (top row) and [Mg/Mn]-[Al/Fe] (bottom row) for the thick disc (left column) and the thin disc (right column).
          }
      \label{fig:simple_line}
\end{figure*}

The thin and thick disc populations show a clear separation in the [$\alpha $/Fe]-[Fe/H] plane \citep{fuhrmann_nearby_1998, bensby_elemental_2003}. Similar to the [Mg/Mn]-[Al/Fe] method, this separation is based on the fact that SNe\,II and SNe\,Ia occur at different timescales and produce different amounts of different elements. Thick disc stars have higher [$\alpha$/Fe]-values than thin disc stars because they are more enriched by SNe\,II, which yield more $\alpha$-elements than iron, than SNe\,Ia, which produce no $\alpha$-elements and more iron.

We used magnesium to represent $ \alpha $ since it gives a clear distinction between the two discs when using the APOGEE DR17 data. A dividing line was manually chosen in the [Mg/Fe]-[Fe/H] plane\footnote{Placing the dividing line with a copula in the [$\alpha$/Fe]-[Fe/H] plane \citep{patil_decoding_2024} is investigated in Appendix \ref{appendix:chem_copula}.}. All the stars below the line were assumed to be thin disc (low-$\alpha$) stars, and those above were assumed to be thick disc (high-$\alpha$) stars. To remove accreted and halo stars, we used a simple cutoff at [Fe/H] $= -1.0$. Stars with metallicities below this limit were assumed to be accreted or halo stars, and those above were assumed to be disc stars.

The results of this selection are shown in Fig.~\ref{fig:simple_line} with the distributions in the [Mg/Fe]-[Fe/H] and [Mg/Mn]-[Al/Fe] planes shown for the thick (Region I) and thin (Region II) discs. The distributions for the two regions are similar to those found in the previous Sect. \ref{section:chem_haywood}. Stars from Region I are densest around ([Mg/Mn] $ \approx 0.4$, [Al/Fe] $ \approx 0.15$) but reach lower values of [Mg/Mn], down to approximately $ -0.1 $. Region II has the densest part of the selection at ([Mg/Mn] $ \approx 0.0$, [Al/Fe] $ \approx 0.0$), and also includes a group of low-Al stars previously identified as halo, located around ([Mg/Mn] $ \approx 0.3$, [Al/Fe] $ \approx -0.2$).

\subsection{Kinematic selection} \label{section:kinematic_method}

\begin{figure}
\centering
\includegraphics[width=0.95\hsize]{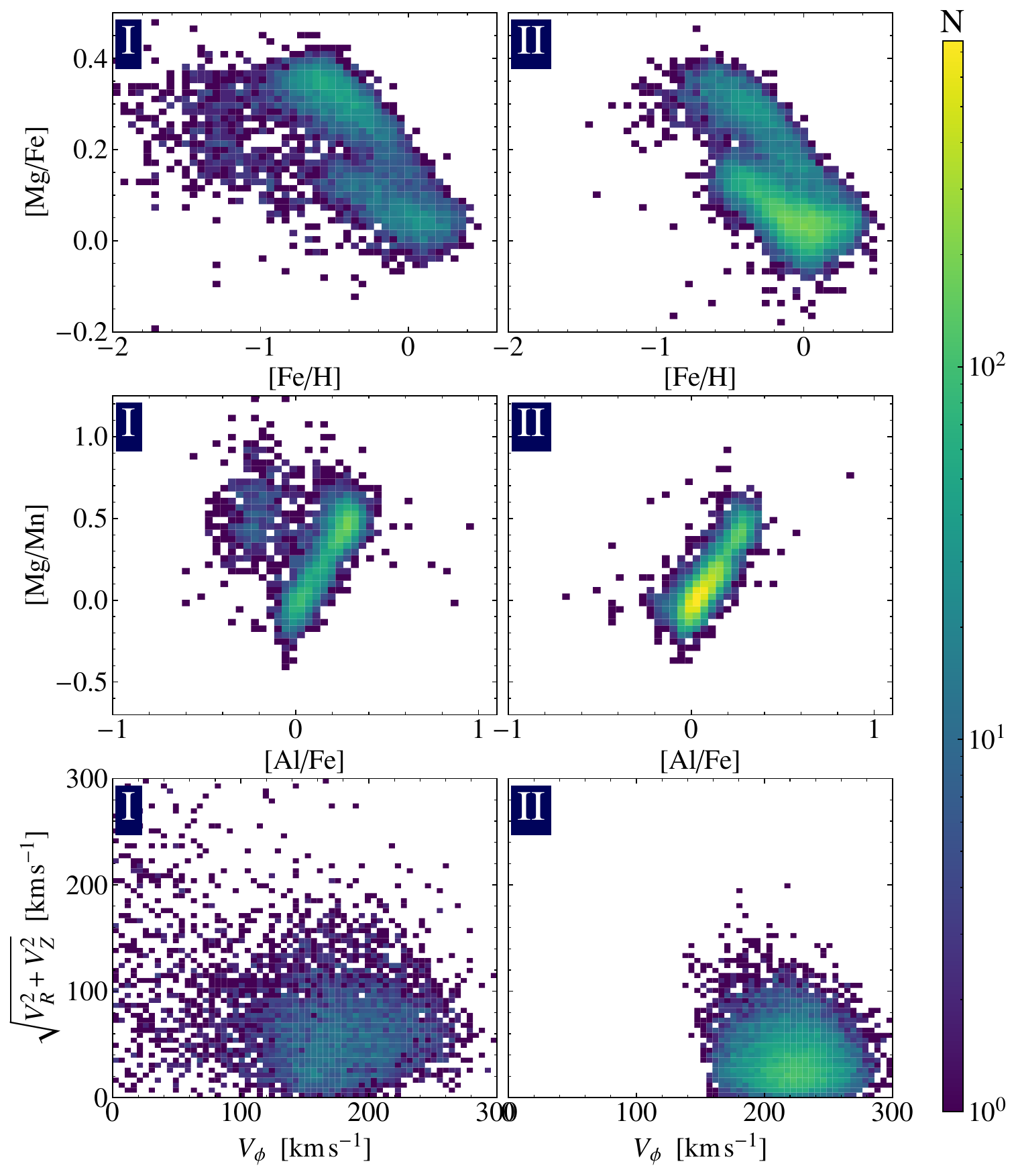}
  \caption{Distribution of stars from the kinematic selection for Region I (\textit{left column}) and Region II (\textit{right column}).
  \textit{Top row}: [Mg/Fe]-[Fe/H]. \textit{Middle row}: [Mg/Mn]-[Al/Fe]. \textit{Bottom row}: Toomre diagrams showing the velocity distributions.
          }
     \label{fig:kinematic results}
\end{figure}

\begin{figure*}
\centering
\includegraphics[width=0.85\hsize]{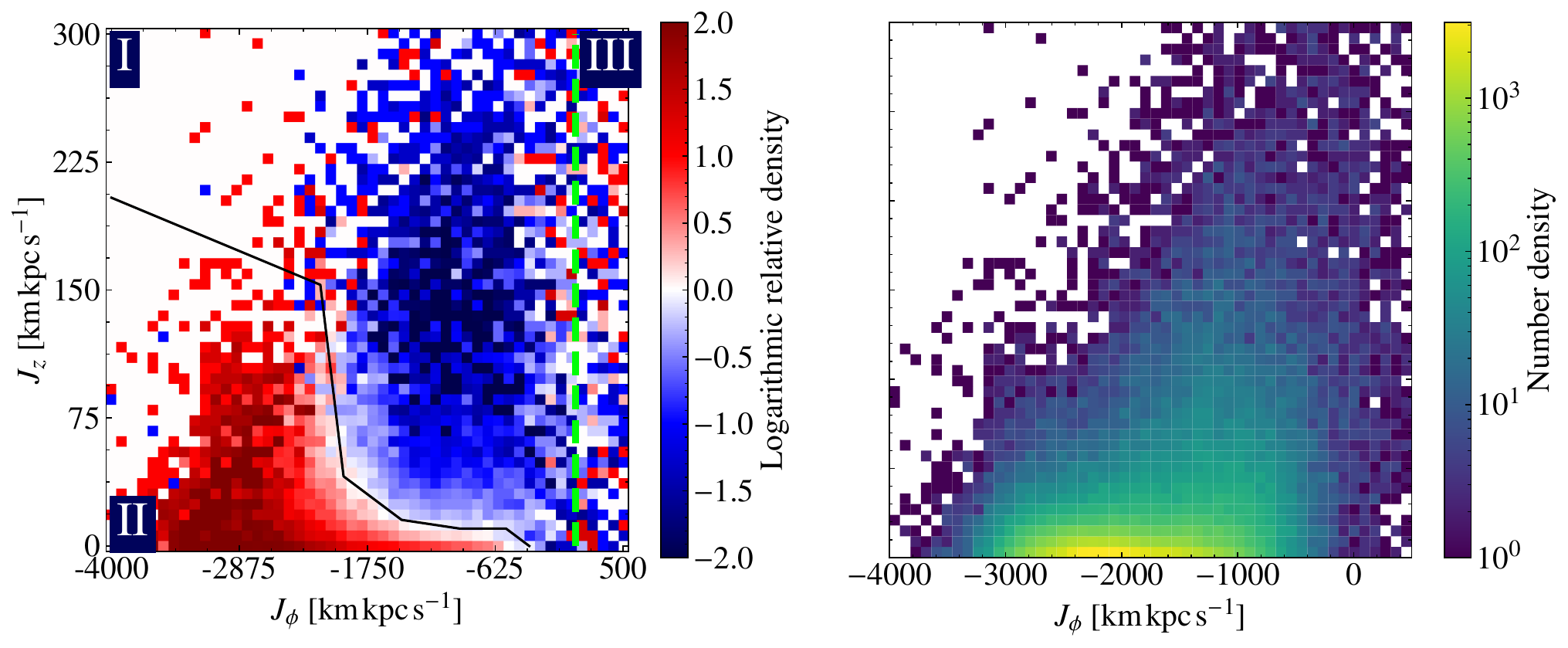}
\caption{
    $J_\phi $-$ J_Z$ plane of the sample.
        \textit{Left}: Relative density of the chemically defined high- and low-$\alpha$ populations in the $J_\phi $-$ J_Z$ plane.
        The colour of each pixel is logarithmic relative density as defined by $\log(N_{thin}) - \log(N_{thick})$, where red means thin-disc-dominated and blue means thick-disc-dominated.
        The two discs are separated by a black line where $ \log(N_{thin}) - \log(N_{thick}) = 0 $.
The dashed green line at $J_\phi = 0$ delineates Region III.
        \textit{Right}: Number density of the sample.
                }
        \label{fig:dynamical}
\end{figure*}

\begin{figure}
\includegraphics[width=0.95\hsize]{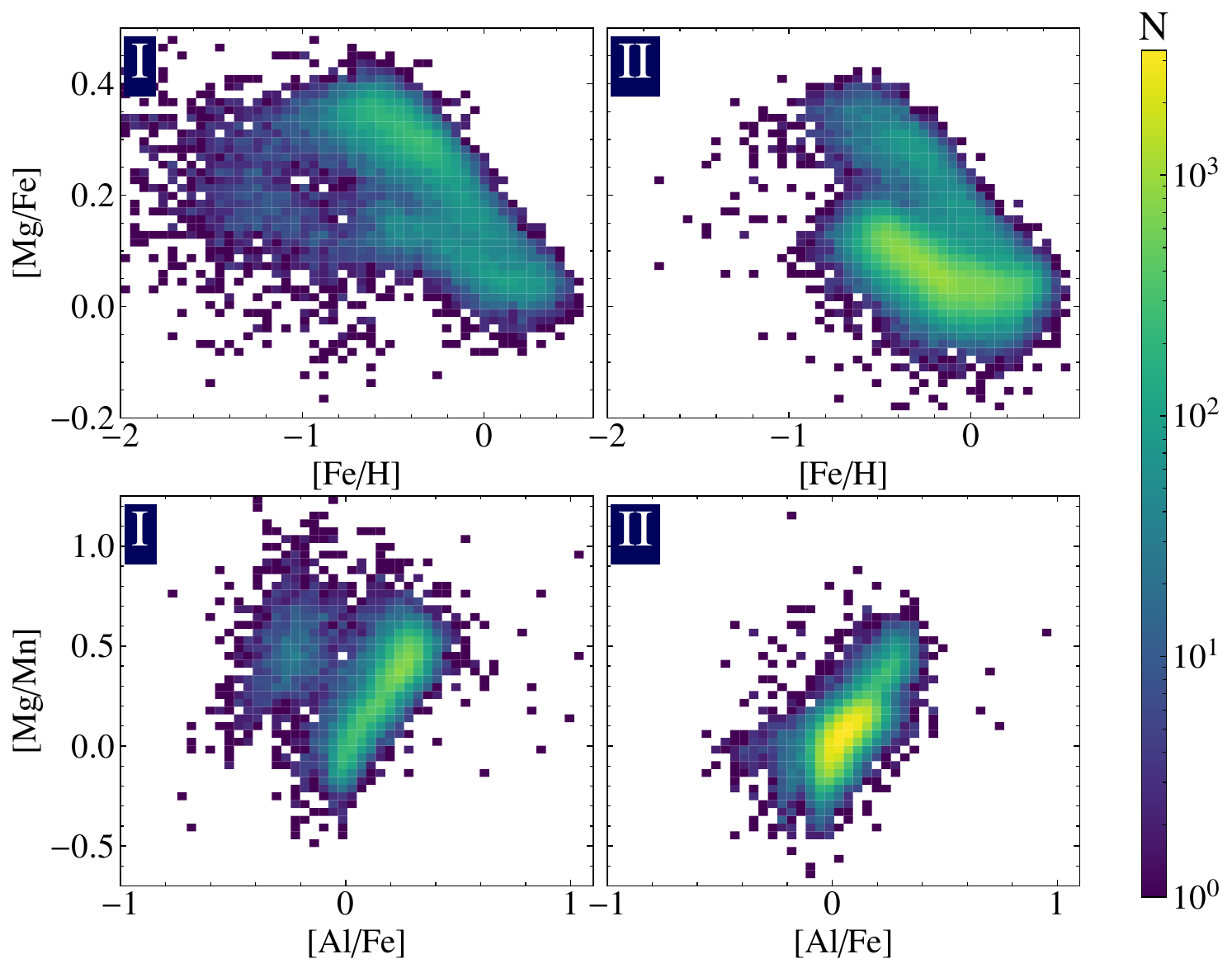}
\caption{Distribution of stars from the dynamical selection.
  \textit{Top row}: [Mg/Fe]-[Fe/H] plane. \textit{Bottom row}: [Mg/Mn]-[Al/Fe].
  Region I shows the thick disc selection and  Region II the thin disc.
          }
     \label{fig:J_phi_z}
\end{figure}

The kinematic method uses the fact that populations of stars retain kinematic imprints of their formation histories.
Thick disc stars move on hotter, more vertically extended orbits and lag behind the local standard of rest in rotational velocity, whereas thin disc stars follow colder, more circular paths. By quantifying these differences statistically, it is possible to assign each star a likelihood of belonging to one component or another, independent of chemical composition.

Following the method defined in \cite{bensby_elemental_2003}, we calculated the relative probability of each star belonging to the thick disc relative to the thin disc, or the thick disc relative to the halo, based on their kinematics. The method approximated the velocities of each Galactic component as drawn from a Gaussian distribution, and that each component occupied a certain fraction of the stellar density in the Galactic mid-plane.
\cite{bensby_elemental_2003} explicitly stated that the values listed were valid for the solar neighbourhood at the Galactic mid-plane.
We therefore restricted the analysis of this model to an area that includes the solar neighbourhood with restrictions on radial distance and azimuthal angle.
It consisted of an annular section of $ 7 $ kpc $ < R_{\mathrm{gal}} < 9 $ kpc, covering $60^{\circ}$, centred on the Sun, with no restriction in height.

Figure~\ref{fig:kinematic results} shows the results of the kinematic selection in the [Mg/Fe]-[Fe/H] and [Mg/Mn]-[Al/Fe] planes. Compared to the chemical selection methods in Sects.~\ref{section:chem_haywood} and \ref{section:chem_simple}, we see that the thick disc mostly contains stars from the high-$ \alpha $ sequence but also stars from the low-$ \alpha $ sequence, and stars with lower metallicities ([Fe/H] $ < -1 $) or low [Al/Fe] stars that were classified as halo in the sections above.
The thin disc mostly contains stars from the low-$ \alpha $ sequence, but a clear signal of the high-$ \alpha $ sequence can also be seen.

\subsection{Dynamical selection} \label{section:dynamic_method}

With the wealth of astrometric data from \textit{Gaia}, one can hope to categorise stars based on their dynamical properties, removing the reliance on detailed elemental abundance data, which are difficult to attain in similar quantities.
There are multiple ways of making a selection using dynamics. The `classical' approach is to model the thin and thick discs as separate quasi-isothermal distribution functions in action space and fit their structural and kinematic parameters simultaneously \citep{binney_models_2011, binney_dynamical_2012}. Authors sometimes mix dynamical and other data, like \cite{helmi_merger_2018}, who use energy and angular momentum data as supporting arguments to argue for their discovery of Gaia-Enceladus, or \cite{mackereth_origin_2019}, who use eccentricity combined with chemical abundances to categorise halo stars.

Our attempt to categorise stars using dynamical properties is done in two steps.
First, we needed to find a suitable action plane for categorising stars into Galactic components, and for this, we turned to azimuthal action ($J_\phi$) and vertical action ($J_Z$). Orbital actions like $ J_\phi $ and $ J_Z$ are adiabatic invariants that quantify the oscillatory motions of a star.
Thin disc stars are typically on near-circular orbits in the Galactic plane, with $J_\phi$ values near that of the local standard of rest, whereas thick disc stars tend to have more eccentric orbits and therefore greater asymmetric drift. This means they have a smaller $ V_\phi $ and thus $J_\phi$ for a given radius \citep{binney_galactic_2008}. $J_Z$ measures the amplitude of vertical oscillations away from the Galactic plane. It is related to the vertical energy and increases with the maximum height reached by a star's orbit. Thin disc stars, which formed in a cold, rotationally supported layer, have small vertical excursions and thus low $J_Z$. Thick disc stars, which are kinematically hotter, whether due to early turbulent formation, heating by mergers, or radial migration, have larger $J_Z$ due to their higher vertical velocity dispersion. This simple picture becomes more complicated if the thin disc is flared. The flared part would consist of thin disc stars in the outer disc with a larger vertical excursion. These stars therefore have high $J_\phi$ and a relatively high $J_Z$. Because $J_\phi$ and $J_Z$ are integrals of motion in an axisymmetric potential, they remain relatively stable over many orbital periods and are less sensitive to the instantaneous phase of a star's motion than to its velocity. The stability of the $J_\phi $-$ J_Z$ plane makes it an effective diagnostic for separating populations with different kinematic temperatures, in theory. In this plane, the thin disc occupies the high-$J_\phi$, low-$J_Z$ area (except for the flare), the thick disc appears at lower $J_\phi$ and higher $J_Z$, and halo stars cluster at low $J_\phi$ with a wide range of $J_Z$ \citep{binney_galactic_2008}.

This leaves the second step: we needed to find a way to divide this plane into areas corresponding to membership in the different Galactic components. To identify the exact locations of the regions for the thin and thick discs, we sorted stars into the high- and low-$\alpha$ components using the abundance-based method described in Sect. \ref{section:chem_simple}, and plotted their relative densities in the $J_\phi $-$ J_Z$ plane in the left panel in Fig. \ref{fig:dynamical}. The white area with $ \log( N_{\rm thin} ) - \log( N_{\rm thick}) = 0 $ is taken as the boundary between the thin and thick disc regions. This leaves the thin disc assigned to the area around $J_Z = 0$ km kpc s$^{-1}$ and at $J_\phi \lesssim -600$ km kpc s$^{-1}$.
The flare of the thin disc is captured as stars with $ J_Z $ up to about 150 km kpc s$^{-1}$ in the area with $J_\phi \lesssim -2000$ km kpc s$^{-1}$. The thick disc is found in the area covering approximately $J_Z \gtrsim 10$ km kpc s$^{-1}$ and $J_\phi \gtrsim -2000$ km kpc s$^{-1}$. To capture non-disc stars, we selected stars with positive $J_\phi$. $J_\phi > 0$ means the stars are on a polar or retrograde orbit and cannot reasonably be said to be members of the disc. The right panel in Fig. \ref{fig:dynamical}, which shows the number density distribution of the sample, reveals no natural overdensities that can be used for separating the sample.

The results of this selection are shown in Fig.~\ref{fig:J_phi_z}. The results are similar to those of the kinematic selection in Sect. \ref{section:kinematic_method}. Region I contains stars from both the high-$ \alpha $ sequence and the high-[Mg/Fe] part of the low-$ \alpha $ sequence. It also contains stars with low [Fe/H] or low [Al/Fe] stars that were classed as accreted by the selections above. Region II shows two sequences, a clear low-$ \alpha $ sequence and a faint high-$ \alpha $ sequence that does not extend below $\rm [Fe/H] \approx-1 $.

\subsection{Selection using age} \label{section:age_method}

\begin{figure*}
\centering
\includegraphics[width=0.45\hsize]{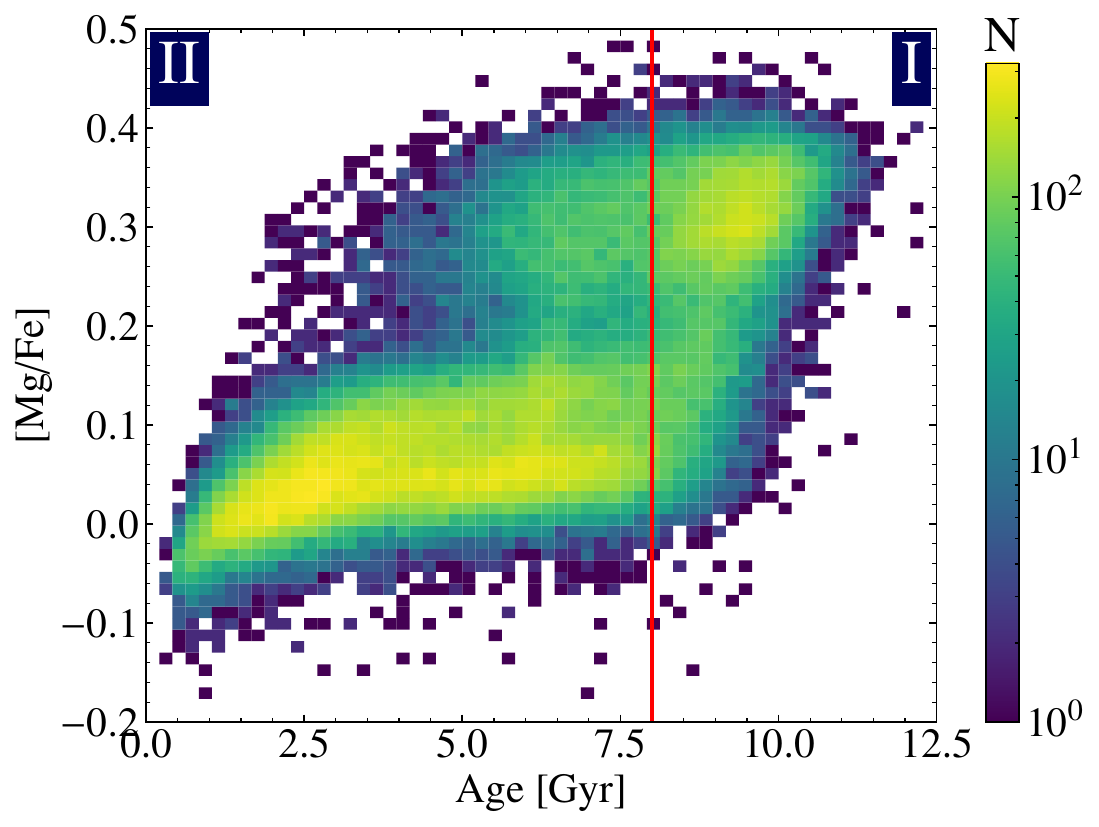}
\includegraphics[width=0.45\hsize]{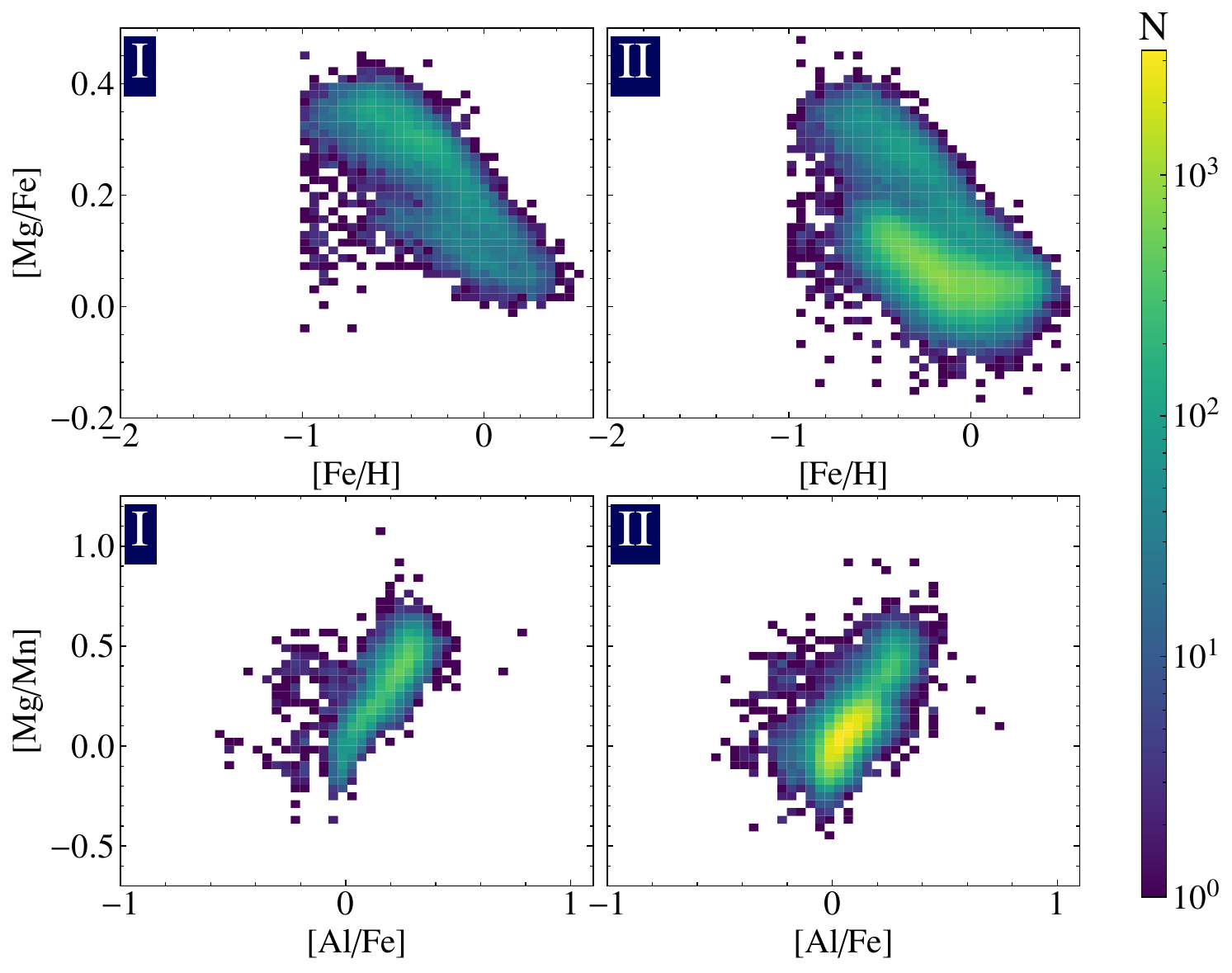}
  \caption{How the age-based method was used to select thin and thick disc stars.
   \textit{Left}: Number density in the [Mg/Fe] - Age plane.
   Region I is the thick disc, and Region II is the thin disc; the two are separated by 8\,Gyr (vertical red line).
   \textit{Right}: Distribution of stars from the selection by age.
   \textit{Top row}:  [Mg/Fe]-[Fe/H]. \textit{Bottom row}: [Mg/Mn]-[Al/Fe].
   \textit{Left column}: Thick disc. \textit{Right column}: Thin disc.
          }
     \label{fig:mg-age}
\end{figure*}

Several of the methods presented above are attempts at using proxies for the ages of the stars. In principle, if the ages can be estimated directly, that should be a very powerful method since it is believed that the majority of thick disc stars formed earlier than the thin disc in the history of the Milky Way \citep{freeman_new_2002}. Most stars do not carry any direct age markers. Instead, age estimates must be inferred from other properties, such as chemical abundances, luminosities, or oscillation frequencies. With the growing availability of high-quality data, it has become possible to construct large-scale catalogues containing stellar age estimates. These provide an opportunity to attempt classification based directly on stellar formation time.

We used the age estimates from the astroNN value-added catalogue for our APOGEE sample. We set a limit on the relative age uncertainty to 50\%, leaving 90\,387 stars (or 85\%) of our sample.
We selected 8\,Gyr as the threshold between the thin and thick disc populations.
This threshold was chosen because \cite{juric_milky_2008}, \cite{bensby_exploring_2014}, and \cite{martig_radial_2016}, among others, indicate that this time is, approximately, after the formation of the thick disc and before the formation of the thin disc.
The left panel of Fig. \ref{fig:mg-age} shows [Mg/Fe] against age for our sample. We see a sequence of low-[Mg/Fe] stars that are young (0--9\,Gyr), and a sequence of high-[Mg/Fe] stars that are mostly older than 4\,Gyr and up to 11\,Gyr, with the densest part around 10\,Gyr. 

Extracting the non-disc sample is not trivial. In Fig.~\ref{fig:mg-age}, we see no visible `tail' of old, non-disc stars on the high-[Mg/Fe] sequence. 
While such a tail should exist, it would not account for all of the accreted stars as they have ages between 8 and 12\,Gyr \citep{das_ages_2020}, which makes a cut in age to extract halo and accreted stars unviable. We therefore used a cut in metallicity to select the accreted population, the same as used in Sect. \ref{section:chem_simple}, with [Fe/H] $< -1.0$ taken to mean a star is not a disc star. The results of this selection are shown in Fig. \ref{fig:mg-age}. In the top row, we see the metallicity cut at [Fe/H] $= -1$, which removed the accreted stars. Region I includes the high-$ \alpha $ sequence and some of the most adjacent parts of the low-$ \alpha $ sequence, while Region II clearly shows the low-$ \alpha $ sequence and a low-density part of the nearby high-$ \alpha $ sequence.

\section{Scale height model} \label{section:methods}

The basis of our analysis is a simple model of the distribution of stars into thin disc, thick disc, and halo as a function of |Z| at different, fixed $ R_{\mathrm{gal}} $.
We used adaptive vertical bins, meaning their positions and widths were assigned such that each contains, if possible, the same number of stars. This led to a total of 25 bins that are densely packed close to the mid-plane, where stellar number density is high, and sparse farther away.
The mid-plane is therefore of greater importance when fitting the model.

We sought to estimate the likelihood that the observed distribution of stars in each bin comes from a given model, which predicts a certain proportion of stars in each of the three components. This is a classic statistical problem of categorising a fixed number of objects into a fixed number of categories, according to some underlying probabilities. The appropriate statistical tool for describing this kind of situation is the multinomial distribution, which allows us to calculate the likelihood of seeing the observed numbers of stars in each category, given the expected proportions. That is, the model predicts that a certain fraction of stars should belong to the thin disc, another to the thick disc, and a final fraction to the halo, with all fractions adding up to one. The multinomial distribution tells us the likelihood of observing the actual star counts we see in a bin, given these predicted fractions. This likelihood function forms the basis for evaluating how well different models explain the observed data.

All the categorisation methods are evaluated by fitting a simple model for the density fraction of each Galactic component as a function of $ Z $ at a given $ R_{\mathrm{gal}} $. The base of the model is an exponential density distribution as a function of height,
\begin{equation}
    N(Z) = \exp \left( - \frac{|Z|}{H} \right)
    \label{eqn:base}
,\end{equation}
where \textit{H} is the scale height.

Two instances of this model ($ N_1 $ and $ N_2 $) are then combined with a constant term for the halo to form the model of the distributions of stars in $ Z $,
\begin{equation}
    N_{\text{total}}(Z) = f_d \, N_1(Z) + (1 - f_d) \, N_2(Z) + C_{halo}
    \label{eqn:total}
,\end{equation}
where $f_d$ is the thick disc fraction in the plane.
We are dealing with relatively small volumes of the Galaxy, and it is therefore acceptable to approximate the halo with a constant term. This simplifies the process and removes a potential source of uncertainty as we do not need to fit an additional halo model. 

The data were split into several radial bins, and for each, the stars were categorised using one of the methods described in Sect.~\ref{section:selections}. We then wanted to determine the model of the Galaxy that best fits this categorisation. Formulated mathematically, this becomes a question of categorising $n$ objects into $k \,(\,=3)$ categories with fixed probabilities. The solution to this problem is the multinomial distribution, and we can express the probability for our model as
\begin{equation}
    P(n_{tn}, n_{tk}, n_{h}|p_{tn}, p_{tk}) = \frac{n!}{n_{tn}!n_{tk}!n_k!} p_{tn}^{n_{tn}}p_{tk}^{n_{tk}}p_{h}^{n_{h}},
    \label{eqn:multinomial1}
\end{equation}
where  $ n $ is the total number of stars, $ n_{tn}, n_{tk} $, and  $ n_{h} $ are the numbers of stars belonging to the thin disc, thick disc, and halo, respectively, and $ p_{tn}, p_{tk} $, and  $ p_{h} $ are the fractions of stars in the thin disc, thick disc, and halo, respectively. To avoid very large factorials, we recast this equation using Gamma functions,
\begin{equation}
    P(n_{tn}, n_{tk}, n_{h}|p_{tn}, p_{tk}) = \frac{\Gamma(n+1)}{\Gamma(n_{tn}+1)\Gamma(n_{tk}+1)\Gamma(n_{h}+1)} p_{tn}^{n_{tn}}p_{tk}^{n_{tk}}p_{h}^{n_{h}}.
    \label{eqn:multinomial2}
\end{equation}
We sought to maximise the product of $P$ for all bins. A convenient way of doing this is to take the logarithm of the probability first, then maximise the results, making the function
\begin{equation}
    \log P(n_{tn}, n_{tk}, n_{h}|p_{tn}, p_{tk}) = \sum_i \log P_i(n_{tn}, n_{tk}, n_{h}|p_{tn}, p_{tk})
    \label{eqn:multinomial4}
,\end{equation}
where the sum is over the vertical bins. We used the \texttt{emcee} \citep{foreman-mackey_emcee_2013} Markov chain Monte Carlo package, which applies the Goodman-Weare algorithm \citep{goodman_ensamble_2010}, to fit the model parameters using Eq.~\ref{eqn:multinomial4} to find the most probable model for each radial bin.

\begin{figure*}
\centering
\includegraphics[width=\hsize]{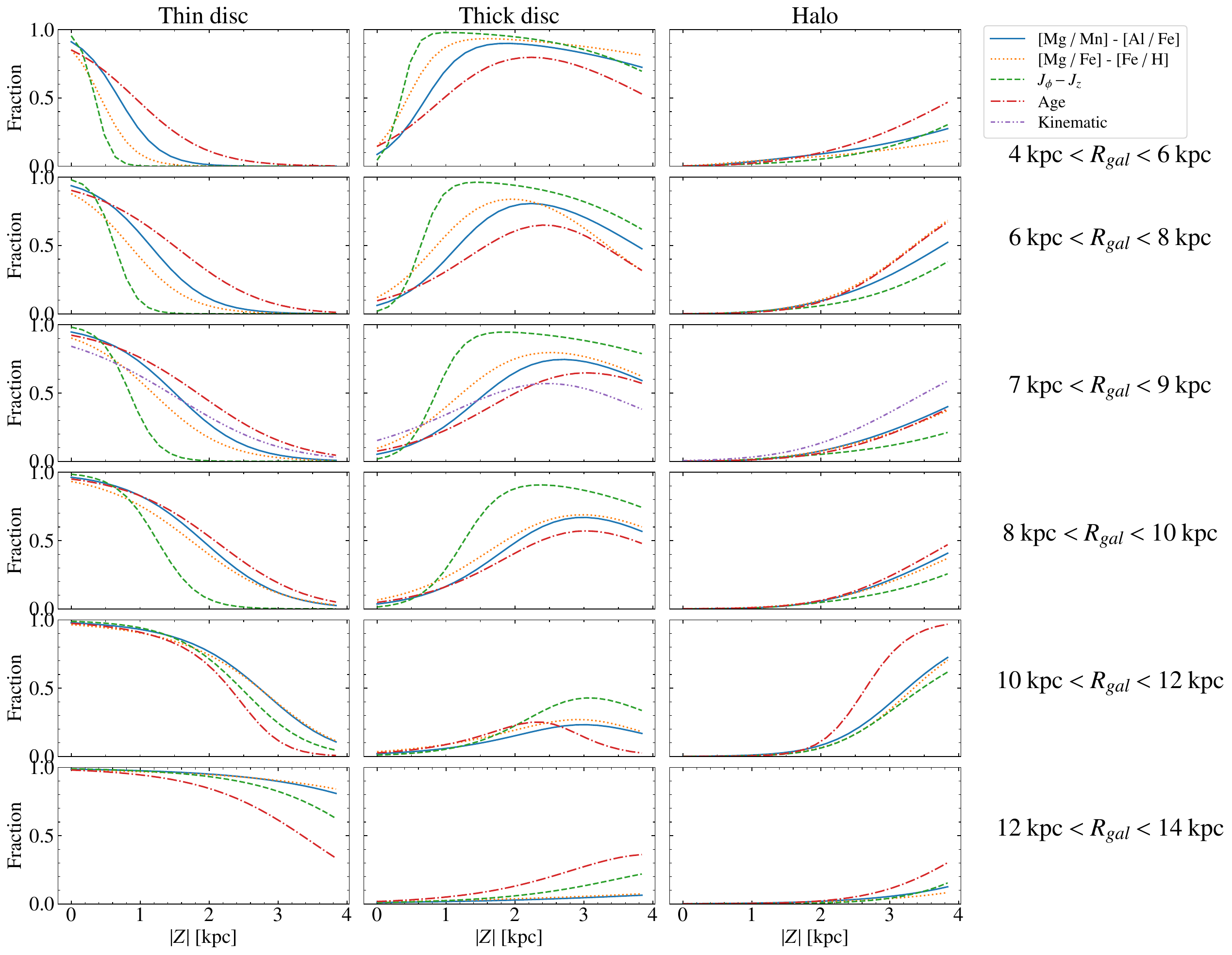}
  \caption{Relative fraction of the thin disc, thick disc, and halo in selected $ R_{\mathrm{gal}} $ bins, based on the methods presented in Sect. \ref{section:selections}. 
  The methods are shown as lines of different colours and styles, as indicated.
The kinematic method is only present in the $ 7 < $ kpc $ R_{\mathrm{gal}} < 9 $ range.
          }
     \label{fig:lines-all-methods}
\end{figure*}

We used our model of the scale heights as functions of $ R_{\mathrm{gal}} $ to derive estimates of the scale lengths of the disc components. By integrating Eq.~\ref{eqn:base} over $Z$, we found the ratio of the surface densities of the thick and thin disc components, $ \Sigma_{tk} $ and $ \Sigma_{tn} $, to be
\begin{equation}
    \frac{\Sigma_{tn}}{\Sigma_{tk}} (R_{\mathrm{gal}}) = \frac{(1 - f_d) H_{tn}(R_{\mathrm{gal}})}{f_d H_{tk}(R_{\mathrm{gal}})}.
    \label{eqn:surf-dense2}
\end{equation}
If we assume both discs are decreasing exponentially with $ R_{\mathrm{gal}} $, we can state the ratio of surface densities of those discs as
\begin{equation}
    \frac{\Sigma_{tn}}{\Sigma_{tk}} (R_{\mathrm{gal}}) \propto e^{R_{\mathrm{gal}} \left( \frac{R_{tn} - R_{tk}}{R_{tn}R_{tk}}\right)},
    \label{eqn:scale-length}
\end{equation}
where $ R_{tn} $ and $ R_{tk} $ are the thin and thick disc scale lengths.

We used bootstrapping in two ways. First, to account for the survey selection function, we treated the inverse of the selection function as a probability density function within the bootstrap procedure. This assigns each star a probability of being drawn in the resampled dataset, which is then used for the rest of the procedure. Second, we used bootstrapping to estimate the uncertainties in our model fits. For each categorisation method, we repeated the categorisation multiple times with a new bootstrapped sample, taking the selection function into account each time. The spread in the resulting density fractions of each component was taken to be their standard deviation, which we used as the fractional uncertainty for each component in the fitting procedure.

\section{Results} \label{section:results}

\subsection{Scale heights}

\begin{table}[t]
\begin{center}
\caption{Model parameters for each method for 2\,kpc wide bins centred on given Galactocentric distances.
\label{tab:results}}
\small
\setlength{\tabcolsep}{4pt}
\begin{tabular}{lcccccc}
\hline\hline
\noalign{\smallskip}
Method 
& \rot{$R=5$ kpc}
& \rot{$R=7$ kpc}
& \rot{$R=8$ kpc}
& \rot{$R=9$ kpc}
& \rot{$R=11$ kpc}
& \rot{$R=13$ kpc} \\
\hline
\noalign{\smallskip}
& \multicolumn{6}{c}{\textit{Scale height thick disc} [kpc]} \\
\noalign{\smallskip}
$\rm [Mg/Mn]$ - $\rm [Al/Fe]$  & 1.4  & 0.83 & 1.0 & 1.0 & 0.88 & 1.7 \\
$\rm [Mg/Fe]$ - $\rm [Fe/H]$   & 1.7  & 0.64 & 1.0 & 1.1 & 0.75 & 1.8 \\
$J_{\phi}$--$J_z$ & 0.89 & 0.81 & 1.2 & 0.99 & 0.96 & 1.1 \\
Ages              & 0.95 & 0.69 & 1.1 & 0.99 & 0.42 & 1.2 \\
Kinematics        & -    & -    & 1.0 & -    & -    & -   \\
\hline
\noalign{\smallskip}
& \multicolumn{6}{c}{\textit{Scale height thin disc} [kpc]} \\
\noalign{\smallskip}
$\rm [Mg/Mn]$ - $\rm [Al/Fe]$  & 0.25 & 0.28 & 0.35 & 0.38 & 0.44 & 0.91 \\
$\rm [Mg/Fe]$ - $\rm [Fe/H]$   & 0.21 & 0.26 & 0.35 & 0.42 & 0.43 & 1.0 \\
$J_{\phi}$--$J_z$ & 0.10 & 0.13 & 0.18 & 0.23 & 0.36 & 0.54 \\
Ages              & 0.34 & 0.34 & 0.44 & 0.42 & 0.27 & 0.52 \\
Kinematics        & -    & -    & 0.48 & -    & -    & -   \\
\hline
\noalign{\smallskip}
& \multicolumn{6}{c}{\textit{Thick disc density fraction in the plane}} \\
\noalign{\smallskip}
$\rm [Mg/Mn]$ - $\rm [Al/Fe]$  & 0.087 & 0.062 & 0.052 & 0.036 & 0.020 & 0.011 \\
$\rm [Mg/Fe]$ - $\rm [Fe/H]$   & 0.14  & 0.12  & 0.095 & 0.066 & 0.035 & 0.017 \\
$J_{\phi}$-$J_z$ & 0.045 & 0.019 & 0.017 & 0.014 & 0.010 & 0.010 \\
Ages              & 0.15  & 0.095 & 0.075 & 0.048 & 0.026 & 0.018 \\
Kinematics        & -     & -     & 0.15  & -     & -     & -     \\
\hline
\end{tabular}
\end{center}
\end{table}

Figure \ref{fig:lines-all-methods} shows how each method fits the thin disc, thick disc, and halo across a range of $ R_{\mathrm{gal}} $. The model parameters are also presented in Table \ref{tab:results}.
Generally, all methods show that the thin disc dominates at large $R_{\mathrm{gal}}$, while the thick disc becomes weak and nearly absent with increasing $R_{\mathrm{gal}}$.

The [Mg/Mn]-[Al/Fe] method and the [Mg/Fe]-[Fe/H] method show similar results. The [Mg/Mn]-[Al/Fe] method finds a slightly higher fraction of thin disc stars at $ R_{\mathrm{gal}} < 9 $ kpc, meaning a correspondingly slightly smaller thick disc fraction in the same regions. This is due to this method including some stars that would usually be included in the high-$\alpha$ sequence in the selection for the thin disc. At higher $ R_{\mathrm{gal}} $, the results are similar.

The dynamical method yields selections with a higher thick disc fraction. By method definition, stars with low vertical action ($ J_Z \leq 10 $ kpc km s$^{-1}$) are classified as thin disc members, except when $J_\phi \approx 0$ kpc km s$^{-1}$. As a result, the disc mid-plane is identified as purely thin disc in all radial bins except the innermost. However, the thin disc density declines more rapidly with $|Z|$ than in the other methods, particularly for $ R_{\mathrm{gal}} < 10 $\,kpc.

The age-based method finds a thin disc fraction that decreases more slowly with |Z| at $ R_{\mathrm{gal}} < 9$ kpc than other methods, despite the thin disc fraction in the plane being lower, and decreases faster than others at $ R_{\mathrm{gal}} > 10 $\,kpc. We also find a larger fraction of halo or accreted stars at $ R_{\mathrm{gal}} > 10 $\,kpc. The reason this method does not have an identical halo fraction to the [Mg/Fe]-[Fe/H] method, despite sharing the definition for halo stars, is that the samples are not identical. The age method only includes stars with reliable age estimates, which is 85\% of the full sample.

The kinematic method is only applied in the $ R_{\mathrm{gal}} = 7 $--$ 9 $\,kpc bin. It finds a lower fraction of thin disc stars in the plane compared to the other methods, and the vertical profiles of both the thin and thick discs are most similar to the age-based method. It finds a large fraction of halo or accreted stars, mostly at the expense of the thick disc, making the thick disc fraction smaller at high |Z|. Finally, it finds that the thin disc fraction does not go to zero by |Z| $ = 4 $\,kpc as for the chemical and dynamical methods, nor stays as large as for the age-based method, but declines almost linearly to just a few per cent at |Z| $ = 4 $\,kpc.

\subsection{Disc parameters}

\begin{figure}
\centering
\includegraphics[width=0.95\hsize]{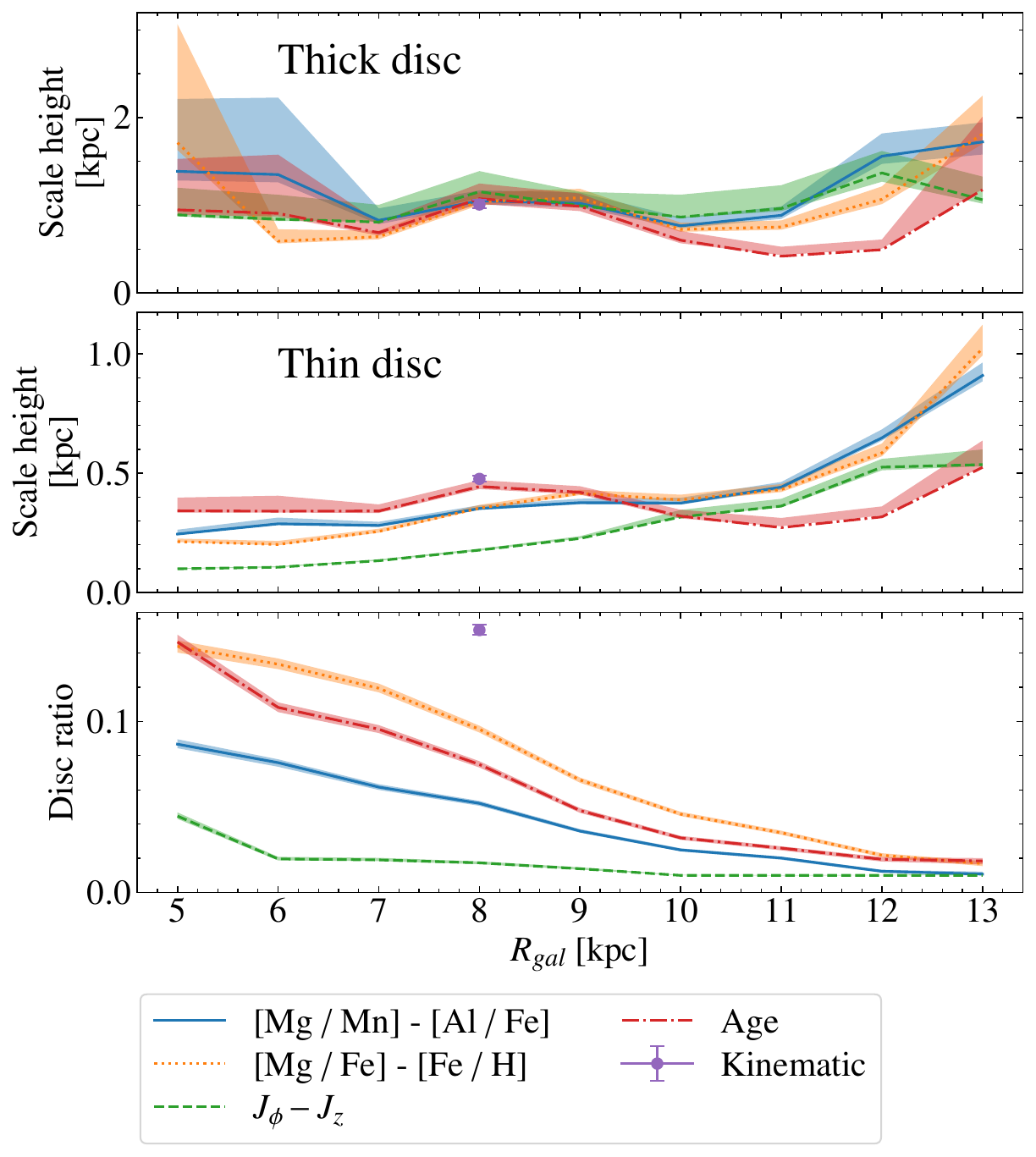}
  \caption{Values of the model parameters for different selection methods over $R_{\rm gal}$.
  The shaded regions are the $ 1\sigma $ uncertainties of the model fit.
  \textit{Top}: Scale height of the thick disc. \textit{Middle}: Scale height of the thin disc. \textit{Bottom}: Thick disc fraction in the plane.
          }
     \label{fig:param-compare}
\end{figure}

Figure \ref{fig:param-compare} shows how the model parameters vary with $ R_{\mathrm{gal}} $ for each method. The thick disc scale height has relatively large uncertainties, but the methods agree on a value of between $ 1.0 $\,kpc and $ 1.25 $\,kpc at $ R_{\mathrm{gal}}\approx 8$\,kpc. 
The thin disc scale height increases with $ R_{\mathrm{gal}} $, showing an exponential increase from between 100\,kpc and 400\,pc at the inner $ R_{\mathrm{gal}} $ edge to between 1\,kpc and 2\,kpc at the outer $ R_{\mathrm{gal}} $ edge. The one exception to this is the age-based method. Its thin disc scale height stays at approximately 400--500\,pc throughout the $ R_{\mathrm{gal}} $ range, though it also shows an increase in the outermost bins, similar to the other methods. The fraction of thick disc stars in the plane (the disc ratio) decreases approximately exponentially with increasing $ R_{\mathrm{gal}} $, starting with values between 4\,\% and 14\,\%, and decreasing to less than 2\,\% in the outer disc at $ R_{\mathrm{gal}} \geq 12$\,kpc.

\begin{figure}
\centering
\resizebox{0.9\hsize}{!}{
\includegraphics{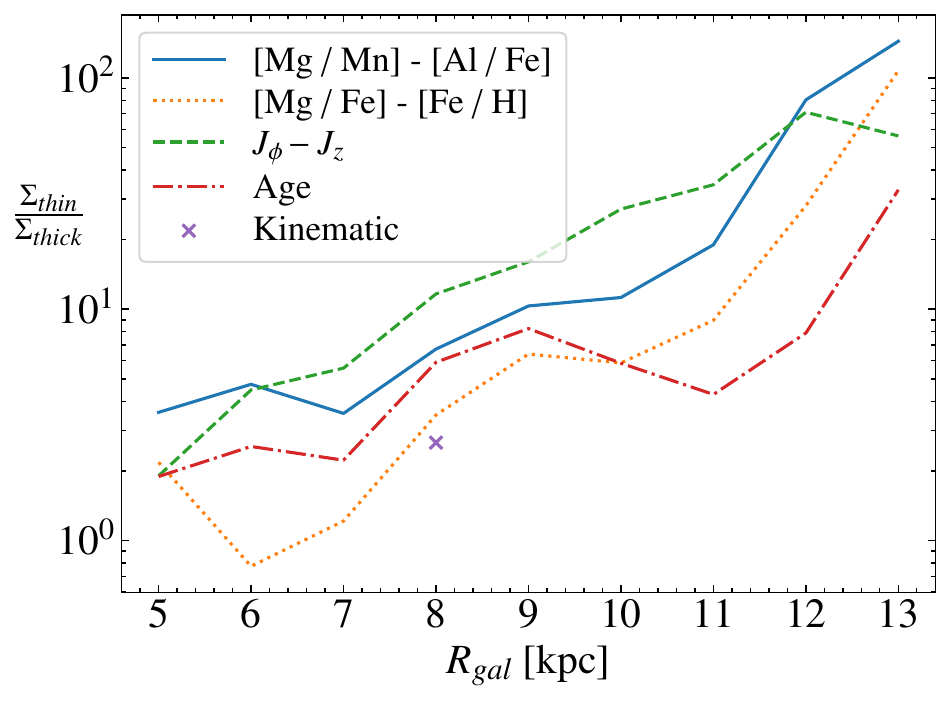}}
\resizebox{0.7\hsize}{!}{
\includegraphics{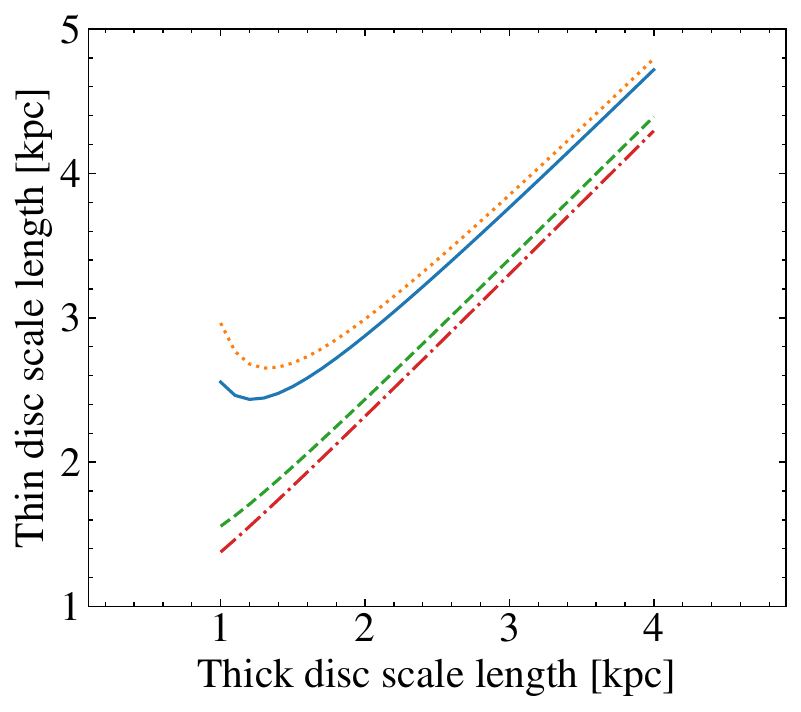}}
  \caption{\textit{Top}: Surface density ratios over $ R_{\mathrm{gal}} $ for the different methods.
  \textit{Bottom}: Best-fit values of the disc scale-lengths for each method.
          }
     \label{fig:scale-length}
\end{figure}
Figure~\ref{fig:scale-length} shows the surface density ratios derived for each method in the top panel. We see that all methods produce approximately exponential curves, indicating that the scale length of the thin disc is larger than the scale length of the thick disc. The lines are offset vertically from each other and mostly parallel, rarely crossing, meaning that the surface density ratios grow at approximately the same pace for all methods. The marker for the kinematic method is slightly below the lines of the other methods, meaning that this method estimates a higher thick disc surface density, which is also seen as a higher disc ratio in Fig. \ref{fig:param-compare}. In the bottom panel in Fig. \ref{fig:scale-length}, we plot the combinations of thin and thick disc scale lengths in Eq. \ref{eqn:scale-length} that produce the best fits to the profiles seen in the top panel. In all cases, the thin disc scale length is larger than the thick disc scale length.

\begin{figure}
\centering
\includegraphics[width=0.95\hsize]{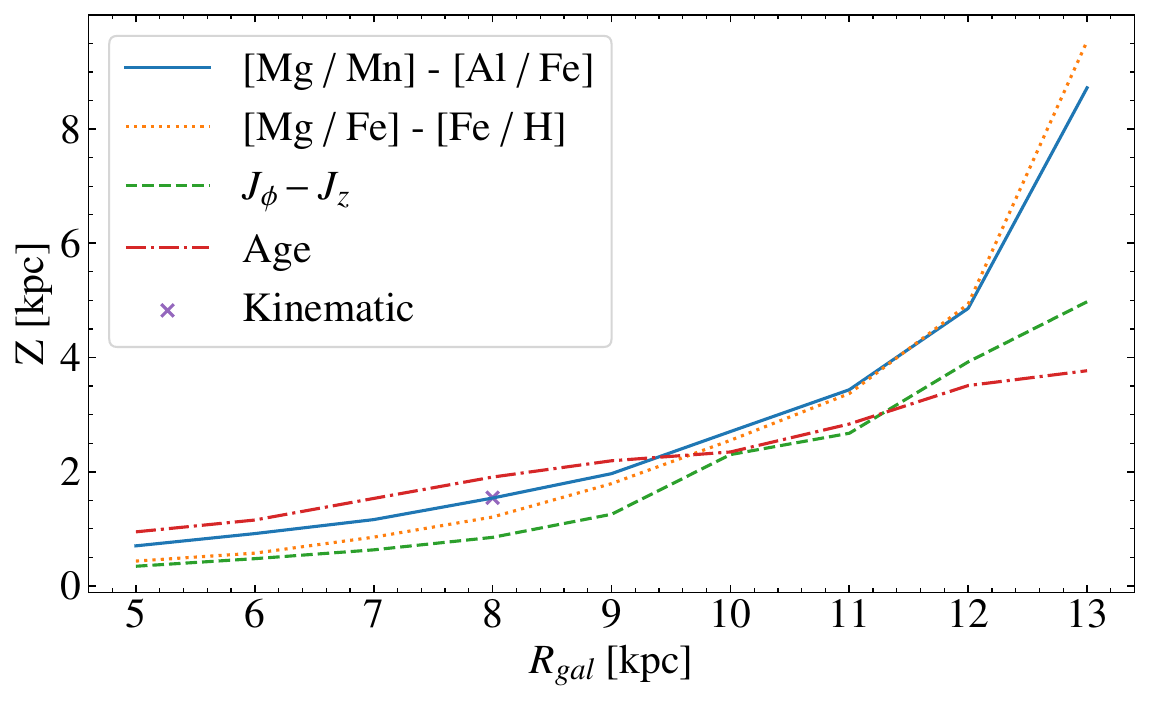}
  \caption{ Z-height where the models for the thick and thin discs have equal density for different selection methods across $ R_{\mathrm{gal}} $.
          }
     \label{fig:break-even}
\end{figure}

Figure~\ref{fig:break-even} shows the height at which the disc goes from being dominated by thin disc stars to being dominated by thick disc stars, that is, they have equal density. All methods show the same general trend of exponential increase, though with some spread in initial values and slopes. Again, the exception is the age-based method, which has the highest initial value and the lowest final value, and shows only a linear increase with $ R_{\mathrm{gal}} $.

\section{Discussion}\label{section:discussion}

The categorisation methods we use in this study are designed to enable comparison between methods, not only to study the properties of the disc components. To achieve this, every star is assigned to a component, and each component is measured as a fraction of the total stellar density. If some stars had been unclassified, the measured fractions would be lower and prevent them from summing to unity.
This requirement sacrifices some sample purity: a method that allows an `uncertainty zone' would produce cleaner samples but would not classify all stars. Our methods lack such zones, unlike many methods that have been applied, which may mean that our results differ slightly.

There are endless ways of comparing the results of the methods.
In Appendix \ref{appendix:results}, we include tables of the number of stars each pair of methods agrees on belonging to each component, and figures showing the distributions of each selection method in all spaces used.
Here, we attempt an overview of these results.
The two chemical methods agree well with each other, as the number of thin disc stars in common between the methods is almost $90\%$ of the number of stars found by either method, and for the thick disc stars it is $68.2 \%$.
The two methods show similar, but not identical, distributions in the chemical spaces, as seen in the right panels in Figs. \ref{fig:mgmnalfe} and \ref{fig:simple_line}, which is what we would expect given the large degree of overlap in selections.
The agreement continues in other spaces as well, with the results of the two methods occupying very similar areas in age, the $J_\phi$-$J_z$ plane, and kinematic space.
That the selections are largely the same shows that the two abundance-based methods produce similar selections despite working in different chemical spaces.

Most of the disagreement between the chemical methods comes from the selection of the thick disc.
There are 8049 stars selected as thick disc by the [Mg/Fe]-[Fe/H] method, but not selected as thick disc stars by the [Mg/Mn]-[Al/Fe] method.
These stars are in the high-$\alpha$ sequence but also have [Mg/Mn] $ < 0.3 $. Almost all are located at $ R_{\mathrm{gal}}< 8$\,kpc within 2\,kpc of the midplane. They show thin disc-like dynamics by being mostly in $J_z < 100 $ kpc km s$^{-1}$, and thin disc-like kinematics by being tightly centred around $V_\phi = 200$ km s$^{-1}$.
The age distribution shows two denser regions, one at 6.5\,Gyr and the other at 9\,Gyr.
A full comparison can be seen in Fig. \ref{fig:special0}.

A total of 3087 stars are classified as thin disc stars by the kinematic method and, at the same time, classified as high-$\alpha$ (thick disc) stars by the [Mg/Fe]-[Fe/H] method, $ 11.5 \% $ of the high-$\alpha$ sample.
These have a similar age distribution to other high-alpha stars, generally between 5 and 11\,Gyr with a concentration at 9.5\,Gyr, but have narrower distributions in $J_\phi$ and $V_\phi$ because this is how the thin disc is defined for the kinematic sample.
A full comparison can be seen in Fig. \ref{fig:opposites}.

The results of all methods show some correlation between age and kinematics. The distribution in ages generally has a large overlap with the thick disc, having most stars between 4 and 11\,Gyr, while the thin disc has most stars between 0 and 10\,Gyr.
The thin disc selections, which tend to trend younger, show a more concentrated distribution in the Toomre plots, and the most concentrated part is at higher $V_\phi$ than the thick disc, $V_\phi \approx 220$ km s$^{-1}$ compared to $V_\phi \approx 180$ km s$^{-1}$. Stars selected as halo or accreted are possibly somewhat clearer as all are old (mostly older than 7 Gyr) and all have low $V_\phi$, with the distribution centred around $V_\phi = 0$ km s$^{-1}$.

Of the methods we test, the kinematic method is the most difficult to evaluate due to its being used in only one radial bin. \cite{schonrich_origin_2009} discussed a comparison between a chemical selection and a kinematic one, finding that the kinematic selection is likely to miscategorise stars because the distribution of kinematic properties of thick disc stars overlaps with the distribution of kinematic properties of thin disc stars.
\cite{bovy_milky_2012} further claimed that the Milky Way possesses a single monotonic distribution in the disc and that imposing a boundary between the `thick' and `thin' discs is not motivated by this continuous distribution. These results agree with ours insofar as there is no sharp distinction between the kinematic properties of the populations of the thick and thin discs.

It would be advantageous if it were possible to categorise stars based on dynamics only, chiefly based on data provided by the \textit{Gaia} mission.
This would remove the need for labour-intensive spectroscopic data collection, but introduce a dependence on the selection of a model of the Galaxy, trading a need for data for a potential bias.
It is therefore noteworthy that our dynamical method produces results that stand out from the others in the inner disc, as seen in the top four rows in Fig. \ref{fig:lines-all-methods}.
This is the result of difficulties in separating well-mixed populations. These mixed populations are also seen by \cite{binney_chemodynamical_2024} as they explore several different areas in chemodynamical spaces for the purpose of constructing a chemodynamical model of the Milky Way. Here, we consider their results through the lens of finding groups analogous to the thin and thick discs in these spaces. For example, in their Fig.~6, they show several plots of the [Mg/Fe]-[Fe/H] plane coloured by mean action ($J_\phi$, $J_Z$, or $J_R$), where each plot contains stars in a range of $J_Z$ and $J_\phi$. Many of the plots show a smooth variation in the measured action across the sample in chemical space, similar to the right panel in our Fig.~\ref{fig:dynamical}. This would make a complete categorisation based purely on dynamics challenging, which is in line with what we find.

The age-based method shows a different result for the $R_{\mathrm{gal}}$-Z shape of the thin disc compared to the other selection methods.
This can be seen in Fig.~\ref{fig:break-even}, where the curve of the age-based method increases approximately linearly with $ R_{\mathrm{gal}} $, while the other methods are approximately exponential.
This is consistent with there being an age gradient in the discs, which would make the inner parts have a larger fraction of younger stars than the outer disc.
This is in line with the age gradients discussed by \cite{martig_radial_2016}. The [Mg/Fe]-age structure in Fig.~\ref{fig:mg-age} shows several interesting properties, which are beyond the scope of this paper to investigate but still worth mentioning. Most obvious is that the high-$\alpha$ sequence is older than the low-$\alpha$ sequence. The low-$\alpha$ sequence also seems to be split in two, with an upper part plateauing at [Mg/Fe] $ \approx 0.12 $ at $ > 3.5 $\,Gyr, while the lower one plateaus at [Mg/Fe] $ \approx 0.5 $ at the same age. The high-$\alpha$ sequence shows two `clumps' of higher density, one at approximately $ 6.5 $ Gyr and the other at approximately $ 9.5 $\,Gyr, which \cite{cerqui_stragglers_2023} associate with stars incorrectly identified as being in the red clump by astroNN.
There is a lower density region with ages $ \lesssim 5 $\,Gyr, which matches the young $ \alpha $-rich stragglers noted by \cite{haywood_age_2013} and further discussed by \cite{cerqui_stragglers_2023}.
For this study, we conclude that there is no one vertical line that can be used to separate the high- and low-$\alpha$ sequences with a high degree of purity. It seems to be possible to get a relatively clean sample of thin disc stars this way, by cutting at 5\,Gyr, but that will lead to a significantly incomplete sample that is still contaminated with a small population of stragglers.

\cite{imig_galactic_2025} conduct an investigation similar to ours. They use red giants from APOGEE and a method similar to ours to account for the selection function to get a representative sample. They combine measurements in age, metallicity, and $ \alpha $ abundance to fit a model of the Galaxy and derive the scale lengths and scale heights for the thin and thick discs, as well as the shape of their profiles. They recover a short scale-length, large scale-height, high-$ \alpha $ component and a larger scale-length, shorter scale-height, low-$ \alpha $ one. The scale heights they derive increase with $ R_{\mathrm{gal}} $, metallicity, and age, and the scale lengths depend on whether the measurement is weighted by mass or light. They find a scale length for the low-$ \alpha $ component of either 2.74 or 3.32\,kpc, depending on weighing, and a scale length of 1.58 or 1.59\,kpc for the high-$ \alpha $ component.
The largest scale heights are in the most metal-rich populations in the inner disc, and the shortest scale heights are in the older, metal-poor populations in the outer disc, as shown in their figures 10 and 11. The scale heights for the low-$ \alpha $ component range from about $ 100 $ pc to about $ 1 $ kpc, and for the high-$ \alpha $ component, the range goes from about $ 400 $ pc to greater than $ 1.5 $\,kpc, which the high-$ \alpha $ component showing a stronger flaring than the low-$ \alpha $ component. This seems to be in tension with our results, which show no noticeable flaring in the thick disc, as well as those of, for example, \cite{bovy_stellar_2016} or \cite{mackereth_age-metallicity_2017}, but is in line with the model proposed by \cite{minchev_formation_2015}.

\section{Summary and conclusions}\label{section:conclusions}

Using red giant stars from APOGEE DR17 and astroNN, we compare five different methods for categorising stars as belonging to the thick or thin discs. These methods are:
\begin{enumerate}
\item separating two populations in the [Mg/Mn]-[Al/Fe] plane,
\item separating two populations in the [$ \alpha $/Fe]-[Fe/H] plane,
\item selecting areas in the $J_\phi $-$ J_Z$ plane,
\item making a cut in estimated ages, and\item making a statistical measure based on velocity dispersions.
\end{enumerate}

Each method produces the relative densities of each component as a function of vertical distance from the Galactic mid-plane at different Galactocentric radii. These results of the different methods were fitted to a simple 1D model of the Milky Way and compared. All methods broadly recover a similar structure of the Galactic disc: the thick disc dominates closer to the Galactic centre, while the thin disc becomes more dominant with $ R_{\mathrm{gal}} $ and shows significant flaring. The fraction of thick disc stars in the mid-plane decreases with Galactocentric radius, reaching a minimum of a few per cent at $ R_{\mathrm{gal}} \geq 12 $\,kpc.

All methods find that the thin disc has a longer scale length than the thick disc.
Based on the relationship between the discs, a thick disc scale length of 2.0\,kpc, as found by \cite{bensby_first_2011} and \cite{bovy_spatial_2012}, corresponds to a thin disc scale length of between $2.3$ and $3.0$\,kpc, depending on the method. A longer thick disc scale length of 3.6\,kpc, as used by \cite{juric_milky_2008}, would yield a thin disc scale length between $3.9$ and $4.4$\,kpc.

No two methods fully agree on the classification of all stars because the stellar populations in the Galaxy are fairly well mixed, particularly in velocity and action space.
The method using the [Mg/Mn]-[Al/Fe] plane is able to separate the disc populations from the old accreted or unevolved populations, but not the discs from each other, while the method using the [$ \alpha $/Fe]-[Fe/H] plane can separate the high-$ \alpha $ population from the low-$ \alpha $ population well, but is less able to separate out the accreted population. Our dynamical method is faced with well-mixed populations and categorises a large number of stars as belonging to the thick disc. The age-based method encounters the issue of overlapping ages for the oldest parts of the thin disc and the youngest part of the thick disc, which presents a formidable obstacle for selecting a clean sample by age alone, so our attempt categorises more stars than the other methods as thin disc stars, particularly in the inner disc. We only applied the kinematic method to the radial bin that includes the solar neighbourhood because this is the only region where this method is valid, and there it encounters the same mixed populations as the dynamical method.

Our results show that the choice of selection method can influence the derived properties of the Milky Way's discs. Abundance-based selections provide higher purity but require spectroscopic data, while kinematic or dynamical methods are more widely applicable but less precise. A superior selection could be achieved by combining multiple methods \citep[e.g.][]{neitzel_dissecting_2025}. The opportunities will be greatly expanded in the near future when the precise astrometric data from \textit{Gaia} is combined with the spectroscopic data provided by 4MOST \citep{de_jong_4most_2019}, in particular the high- and low-resolution surveys of the Milky Way bulge and disc \citep{bensby_4most_2019, chiappini_4most_2019}, and WEAVE \citep{jin_wide-field_2024}.

\begin{acknowledgements} PM gratefully acknowledges support from a project grant from the Swedish Research Council (Vetenskapr\aa det, Reg: 2021-04153). TB and SA acknowledge support from project grant No.~2018-04857 from the Swedish Research Council. Some of the computations in this project were completed on computing equipment bought with a grant from The Royal Physiographic Society in Lund.
    This work has made use of data from the European Space Agency (ESA) mission
    {\it Gaia} (\url{https://www.cosmos.esa.int/gaia}), processed by the {\it Gaia}
    Data Processing and Analysis Consortium (DPAC,
    \url{https://www.cosmos.esa.int/web/gaia/dpac/consortium}). Funding for the DPAC
    has been provided by national institutions, in particular the institutions
    participating in the {\it Gaia} Multilateral Agreement.
    Funding for the Sloan Digital Sky  Survey IV has been provided by
    the Alfred P. Sloan Foundation, the U.S. Department of Energy Office of Science,
    and the Participating Institutions.
    This research has made use of NASA’s Astrophysics Data System.
    This work made use of the following software packages for Python,
    \verb|AstroPy| \citep{astropy_collaboration_astropy_2022}, 
    \verb|emcee| \cite{foreman-mackey_emcee_2013}, 
    \verb|Numpy| \citep{harris_array_2020}, 
    \verb|Matplotlib| \citep{hunter_matplotlib_2007},
    \verb|SciPy| \citep{virtanen_scipy_2020}.
\end{acknowledgements}

\bibliographystyle{aa}
\bibliography{export-bibtex_ads2nty.bib}

\begin{appendix}
\section{Data selection} \label{appendix:data}
To get a sample that is representative of the general stellar population to the greatest extent possible while also containing stars with high-quality spectral data, we used the quality flags provided by APOGEE.
We used both quality flags for each star and element flags for each element.

We required that the flag \texttt{EXTRATARG} = 0, which indicates that the star was randomly selected; see Appendix \ref{appendix:selection}.
We required that the flag \texttt{MEMBERFLAG} = 0 be set to remove any stars that are members of clusters or dwarf galaxies.

Further, we required that the following APOGEE quality flags are not set,
\begin{itemize}
    \item \texttt{BAD\_PIXELS}
    \item \texttt{VERY\_BRIGHT\_NEIGHBOR}
    \item \texttt{STAR\_BAD}
    \item \texttt{CHI2\_BAD}
    \item \texttt{VSINI\_WARN}
    \item \texttt{DUPLICATE}
    \item \texttt{PERSIST\_HIGH}
    \item \texttt{SUSPECT\_BROAD\_LINES}
    \item \texttt{SUSPECT\_ROTATION}
    \item \texttt{M\_H\_BAD}
    \item \texttt{CHI2\_WARN}
\end{itemize}

We made use of APOGEE's element flags for each abundance considered in our study, [Fe/H], [Mg/Fe], [Mn/Fe], and [Al/Fe].
We required that the following flags for each element are not set,
\begin{itemize}
    \item \texttt{GRIDEDGE\_BAD}
    \item \texttt{CALRANGE\_BAD}
    \item \texttt{OTHER\_BAD}
    \item \texttt{PARAM\_MISMATCH\_BAD}
    \item \texttt{TEFF\_CUT}
\end{itemize}
We also required the flag \texttt{FE\_H\_FLAG} = 0, which removes entries with poor [Fe/H] values.

Finally, we used these additional cuts to select only red giants,
\begin{itemize}
    \item teff < 7000 K
    \item logg < 2.8
    \item SNRev > 80
\end{itemize}
SNRev is the revised S/N estimate, which is used instead of SNR because it avoids persistence issues present in some of the detectors. It is recommended to be used in place of SNR by \cite{holtzman_apogee_2018}.

\section{Selection function} \label{appendix:selection}
We used a modified version of the \texttt{apogee\_sf} method from the \texttt{gaiaunlimited} package\footnote{\url{https://github.com/gaia-unlimited/gaiaunlimited}} to compute the selection function for the stars in our sample.
The method for obtaining the APOGEE selection function in \texttt{gaiaunlimited} is described in \cite{cantat-gaudin_uniting_2024}.
This method is based on comparing the number of stars in APOGEE to the number of stars in 2MASS \citep{skrutskie_two_2006} for each field and colour-magnitude bin.
2MASS is assumed to be a representative intermediary sample that can be used for this purpose.
Using this package, we find that 37\% of the stars in our sample were assigned a selection function of exactly zero, meaning that a selection function cannot be described and that the star cannot be used.

We modified \texttt{apogee\_sf} to expand the selection and use stars in fields (named areas on the sky) with mixed designs (selections of stars), as well as fields that were not included in the \texttt{gaiaunlimited} data by using a more relaxed definition of the Main Red Star Sample compared to \cite{cantat-gaudin_uniting_2024}.
In their work, a star is part of the Main Red Star Sample if it has the \texttt{EXTRATARG} = 0, a \texttt{PROGRAMNAME} of \texttt{bulge}, \texttt{disk}, \texttt{disk1}, \texttt{disk2}, or \texttt{apogee}, and a \texttt{FIELD} name that follows the pattern of \texttt{LLL$\pm$BB} where L and B are Galactic longitudes and latitudes.
We relaxed these conditions by only using the \texttt{EXTRATARG} = 0 requirement.

The value of the selection function is computed in two parts: the selection fraction and the extinction correction.
The selection fraction is computed by dividing the number of stars in APOGEE by the number of stars in 2MASS for each region of the colour-magnitude diagram for each field.
The extinction correction is the fraction of 2MASS stars within the magnitude range, ignoring the colour limits, that have extinction values from 2MASS.
The final selection function is computed by multiplying the selection fraction by the extinction correction.
This updated version of \texttt{apogee\_sf} returns a selection function of exactly zero for 15\% of our sample instead of 37\%.

\section{Chemical selection with a copula split} \label{appendix:chem_copula}
\begin{figure*}
\centering
\includegraphics[width=\hsize]{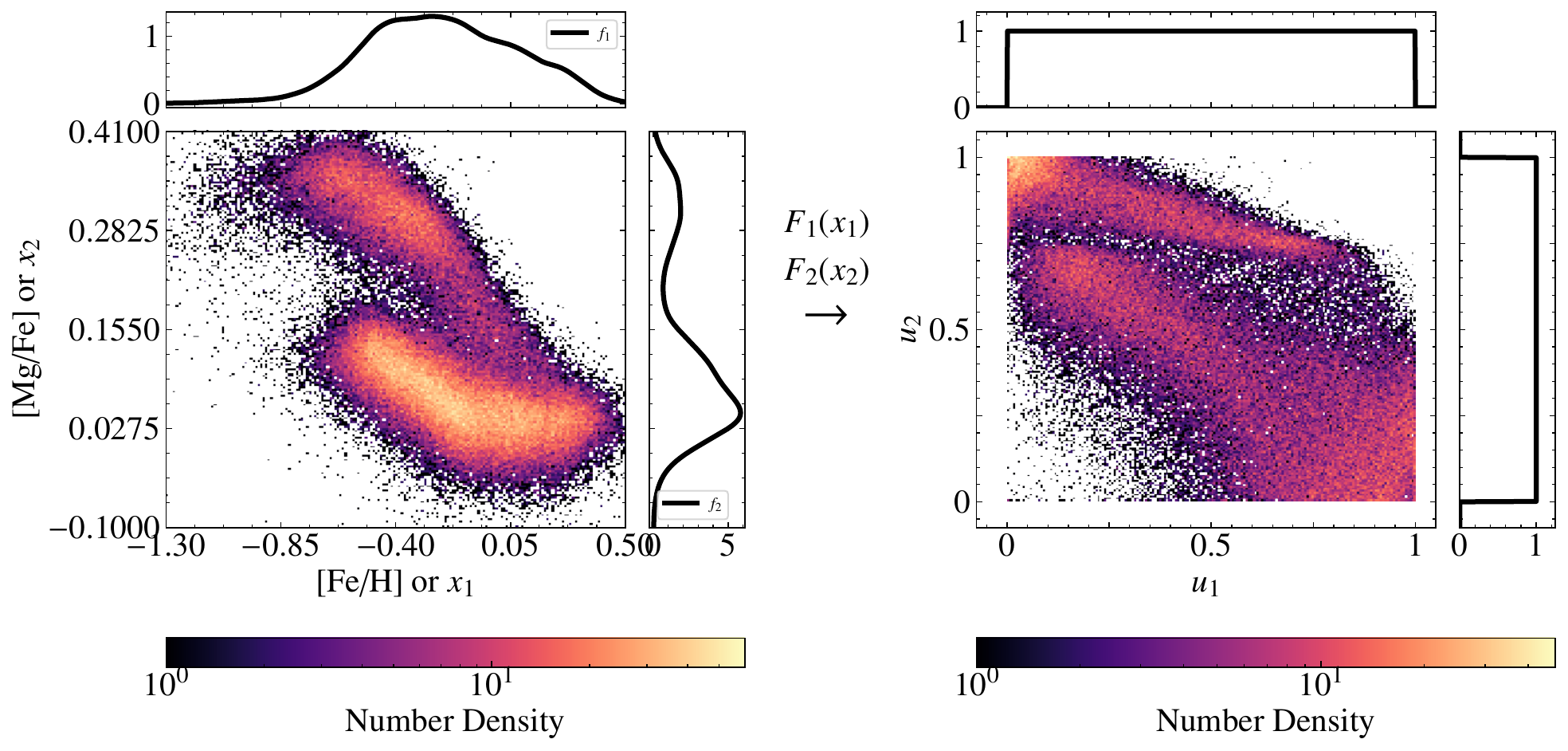}
  \caption{Transformation of the sample in the [Mg/Fe]-[Fe/H] plane into copula space with flat probability distributions.
  Reproduction of Fig. 5 from \cite{patil_decoding_2024}.
          }
     \label{fig:copula transform}
\end{figure*}

\begin{figure}
\centering
\includegraphics[width=\hsize]{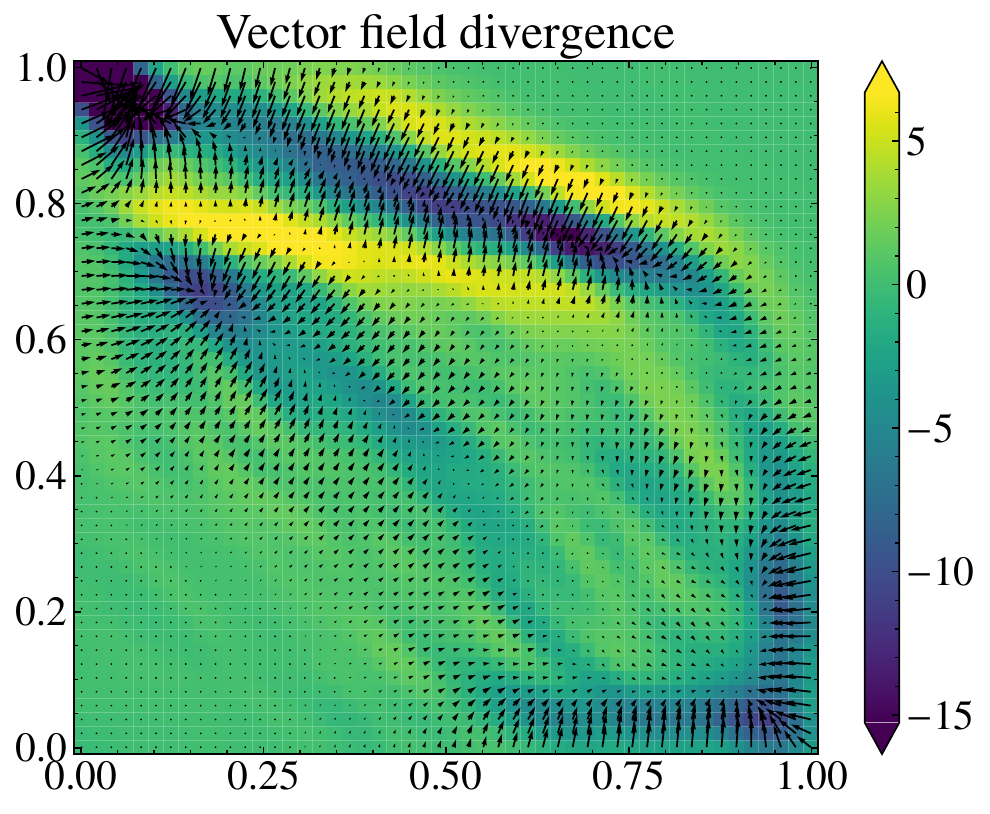}
  \caption{Divergence of the density scalar field of the probability distribution of copula space.
  The vectors show the gradient. The colour shows the divergence.
          }
     \label{fig:vector div}
\end{figure}

\begin{figure}
\centering
\includegraphics[width=\hsize]{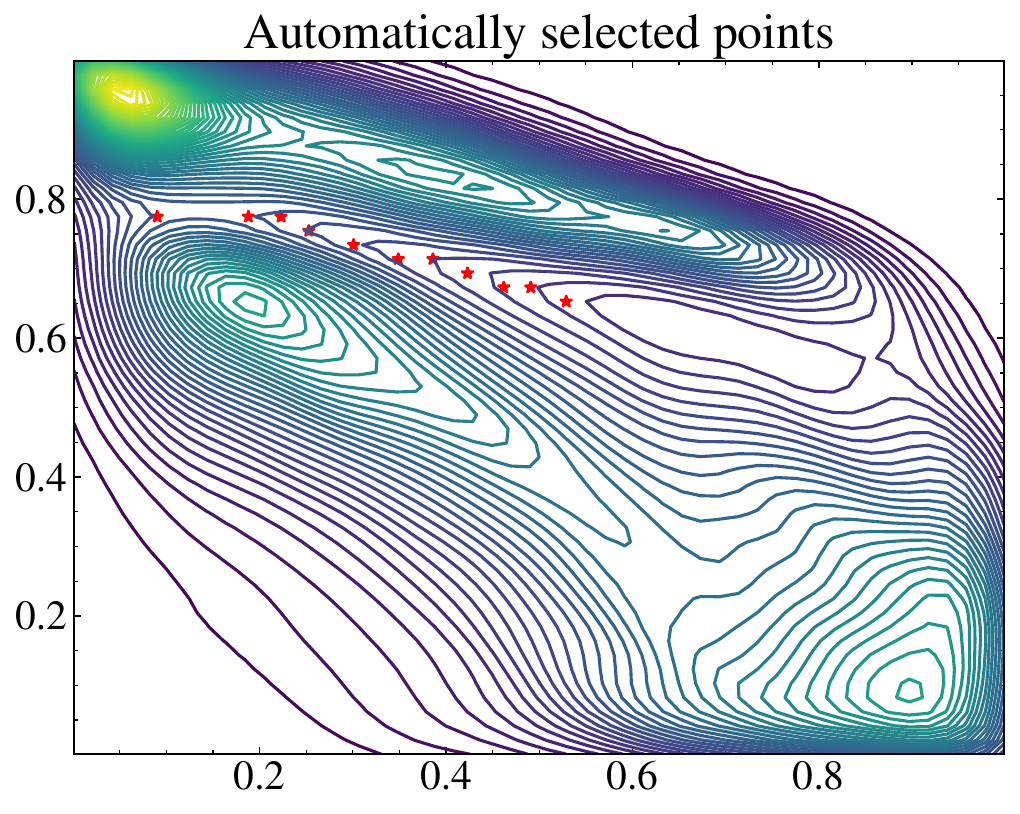}
  \caption{Algorithmically placed points on the contour lines of the gradient of the probability distribution.
          }
     \label{fig:auto points}
\end{figure}

\begin{figure}
\centering
\includegraphics[width=\hsize]{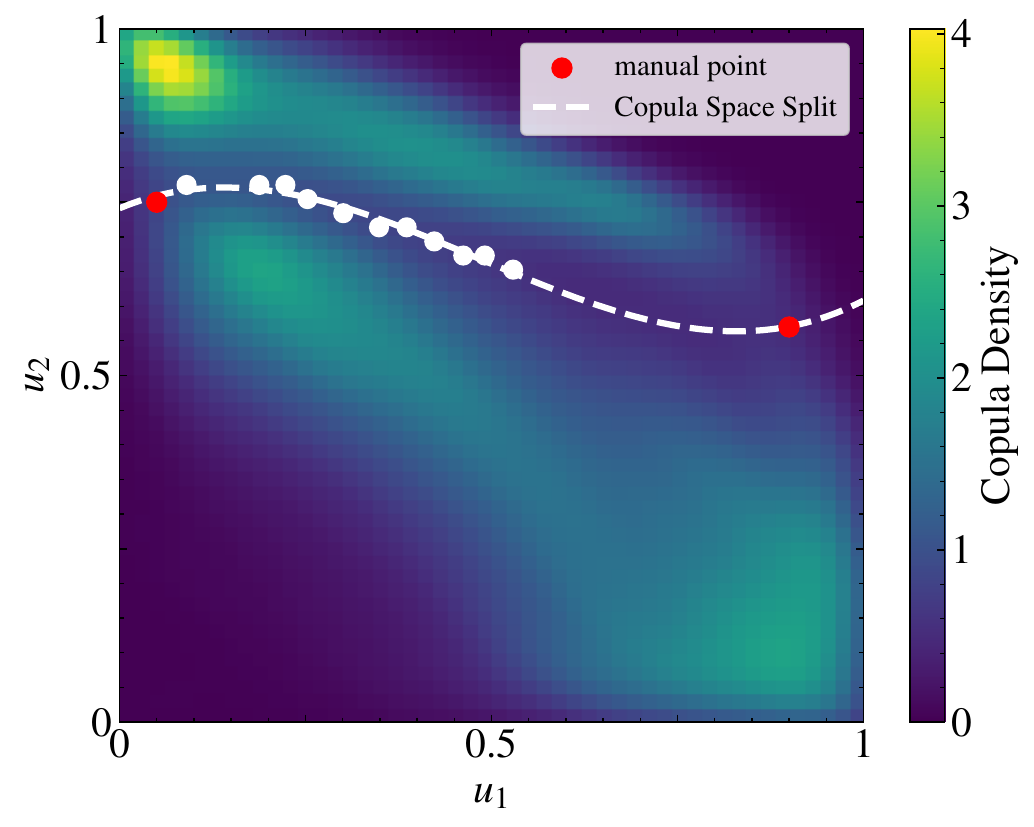}
  \caption{Illustration of our automated procedure for selecting the copula.
        White points are placed by the algorithm. Red points are placed manually. The dashed white line is a third-degree polynomial fitted to all the points.
          }
     \label{fig:copula fit}
\end{figure}

\begin{figure}
\centering
\includegraphics[width=\hsize]{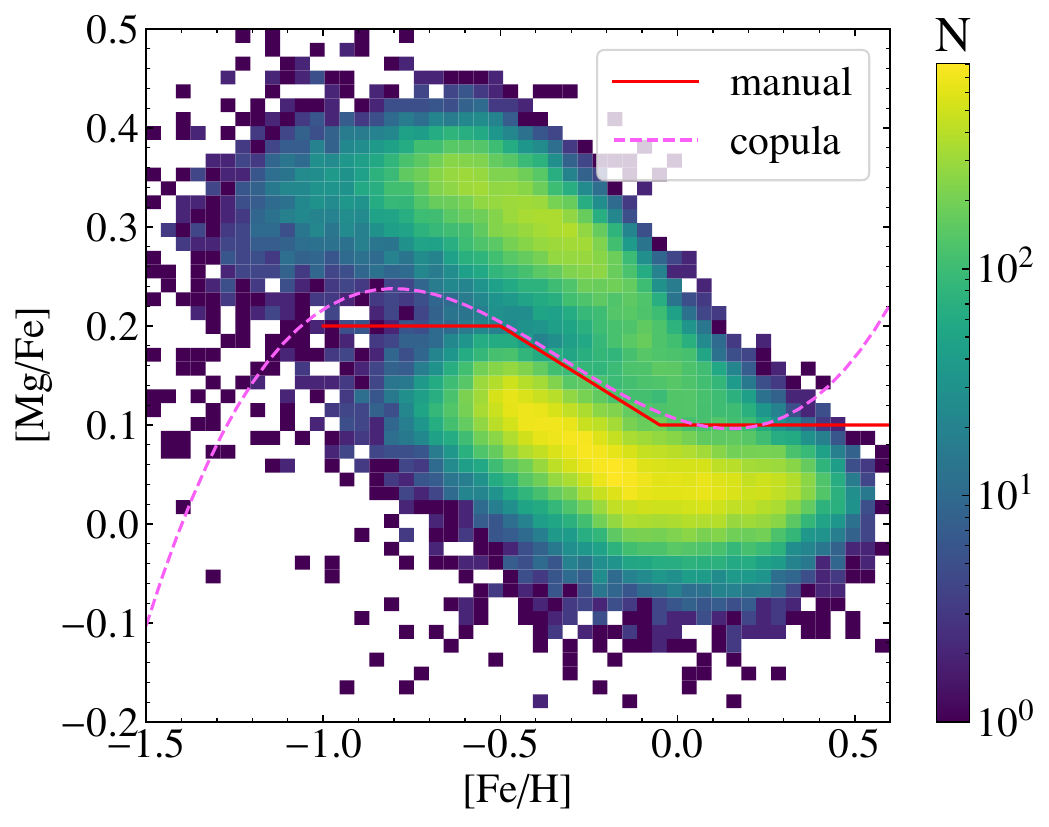}
  \caption{[Mg/Fe]-[Fe/H] plane with both a copula line and the manually drawn line from Sect. \ref{section:chem_simple}.
          }
     \label{fig:both_lines}
\end{figure}

We investigated the method of separating the high- and low-$\alpha$ sequences in the [Mg/Fe]-[Fe/H] plane following \cite{patil_decoding_2024}, who present a method using copulas to select the lowest density regions between the two.
A copula is, in this context, a distribution function for which the marginal probability distribution of each variable is uniform on the interval [0, 1].
\cite{patil_decoding_2024} contains the original description of the algorithm implemented here, as well as the mathematical background for this method\footnote{The original paper \cite{patil_decoding_2024} only contains the description of an algorithm to select the points of a copula, but no implementation.
For this reason, we developed an implementation of the algorithm as part of this project.
The code is available at \url{https://github.com/AlinderS/astro-copula-algorithm}.
}.
In it, the probability distribution functions of the sample along the axes of the [Mg/Fe]-[Fe/H] plane are transformed into flat distributions; see Fig. \ref{fig:copula transform}.
An automated procedure then identifies an area in the transformed plane suitable for splitting the distribution and places a number of points in it.
This procedure considers the probability density space as a scalar field.
The region separating the upper and lower sequences is identified as an area of high divergence in the gradient of this scalar field; see Fig. \ref{fig:vector div}.
Within this region, the algorithm places points at the extreme left or right points of the probability density contour lines, depending on their orientation, as shown in Fig. \ref{fig:auto points}.

A weakness of this implementation of the point placing algorithm is that it is not normally able to find points to place at the edges of the space.
For this reason, a couple of additional points are placed manually in the area where the algorithm cannot identify optimal placements as a way to guide the fitted line to the edges.
These points are placed according to the same criteria as the algorithm uses and are marked in red in Fig. \ref{fig:copula fit}.
A third-degree polynomial is fitted to the points, as it is the simplest polynomial that adequately captures the shape of the separation between the upper and lower regions in probability density.
The polynomial curve is then transformed back to the chemical plane and used to separate the sequences, shown in Fig. \ref{fig:both_lines}, together with the manual line we used in our method described in Sect. \ref{section:chem_simple} for comparison.
It shows that it is possible to closely emulate the selection of a copula using a simple but well-informed line placed manually in the plane.

The argument for using this method is that it allows one to make an unambiguous selection in the [Mg/Fe]-[Fe/H] plane, particularly regarding the placement of the line splitting the high- and low-$\alpha$ sequences in the [Mg/Fe] direction.
We find this to be a useful application of this technique, but, as pointed out by \cite{patil_decoding_2024} themselves, the most efficient way of using it is to use pre-computed values rather than recreating the method, which is also how it is being used in practice by \cite{imig_galactic_2025}.

\section{Method categorisation figures} \label{appendix:results}
To give a clearer idea of where our different samples are located in physical and chemical space, we present here figures showing the results of each method presented in Sect. \ref{section:selections} with distributions in X-Y location, $ R_{\mathrm{gal}} $-Z location, the [Mg/Fe]-[Fe/H] plane, the [Mg/Mn]-[Al/Fe], the $J_\phi$-$J_z$ plane, the Age-[Mg/Fe] plane, and a $ V_\phi $-$\sqrt{V_R^2+V_Z^2}$ Toomre plot for the thin disc, thick disc, and halo categories.
The thin and thick disc categories share a colour scale for easy comparison.
The distributions in the halo category (third row of each figure) are shown with a linear colour scale due to low numbers of stars, unlike the thick and thin discs, which use a logarithmic colour scale.
The y-axis labels are shown on top to make the figures fit.

Tables \ref{tab:thin pairs} and \ref{tab:thick pairs} show counts of how many stars each pair of methods agrees on are in either the thin or thick disc.
The tables list all methods along each axis, making the diagonal elements the total number of stars each method finds for that component, and allow us to investigate how well the selections agree with each other.
The duplicate elements in the upper half of each table have been replaced with the fraction of the number of stars selected by either of the methods.
If we call the pair of methods we consider $ A $ and $ B $, the fraction is $ \frac{A \cap B}{A \cup B} $.

In Table \ref{tab:thin pairs}, we see a high level of similarity between the two chemical methods.
The number of stars selected as thin disc stars by both methods is almost $90\%$ of the number of stars selected by either method.
For the thick disc, in Table \ref{tab:thick pairs}, the agreement is smaller but still good, with $68.2 \%$ of the number of stars selected by either method.
The results of dynamic method agree well with those of the other methods for the thin disc selection, with the common fractions being $84\%$, $83\%$, and $66\%$ for the [Mg/Mn]-[Al/Fe] method, the [Mg/Fe]-[Fe/H] method, and the age-based method, respectively, but not for the thick disc selection, which has common fractions of only $ 50\% $, $ 54\% $, and $ 35\% $ for the same methods.
The age method is similar, with higher levels of agreement for the thin disc, $71\%$, $70\%$, and $66\%$, compared to the thick disc, $44\%$, $51\%$, and $35\%$.
In general, agreement is higher in the thin disc selection.

A more detailed study is possible.
By going beyond just the pairs and looking at stars in common between three, four, or all five methods, it is possible to gain much deeper insight into how each method relates to each other method.
A full study would include 31 different categories per component, consisting of all possible combinations, but such a study is beyond the scope of this paper.

\begin{table}
        \caption{Number of thin disc stars selected in common for each pair of methods.}
        \begin{tabular}{c|ccccc}
                \hline \hline
                Method & \rotatebox{90}{\makecell{[Mg/Mn] \\-[Al/Fe]}} & \rotatebox{90}{\makecell{[Mg/Fe] \\-[Fe/H]}} & \rotatebox{90}{Dynamic} & \rotatebox{90}{Age} & \rotatebox{90}{Kinematic} \\
                \hline
                [Mg/Mn]-[Al/Fe]  & \textbf{84840} & 89.9\% & 84.5\% & 71.4\% & 14.3\% \\
                {[Mg/Fe]-[Fe/H]}  & 76814 & \textbf{77457} & 83.1\% & 70.0\% & 14.0\% \\
                Dynamic  & 75071 & 71056 & \textbf{79117} & 66.1\% & 14.6\% \\
                Age  & 65569 & 61728 & 60366 & \textbf{72511} & 15.5\% \\
                Kinematic  & 12433 & 11291 & 11923 & 11667 & \textbf{14405} \\
                \hline
        \end{tabular}
        \label{tab:thin pairs}
        \tablefoot{Numbers on the diagonal are the total number of stars selected for that method, marked in bold. The upper half contains the fraction of the smaller of the two totals, giving a comparison to the maximum overlap. The kinematic method is applied to a smaller space and therefore includes fewer stars.}
\end{table}

\begin{table}
        \caption{Number of thick disc stars selected in common for each pair of methods.}
        \begin{tabular}{c|ccccc}
                \hline \hline
                Method & \rotatebox{90}{\makecell{[Mg/Mn] \\-[Al/Fe]}} & \rotatebox{90}{\makecell{[Mg/Fe] \\-[Fe/H]}} & \rotatebox{90}{Dynamic} & \rotatebox{90}{Age} & \rotatebox{90}{Kinematic} \\
                \hline
                [Mg/Mn]-[Al/Fe] & \textbf{19382} & 68.2\% & 49.8\% & 44.1\% & 13.6\% \\
                {[Mg/Fe]-[Fe/H]} & 18713 & \textbf{26762} & 54.1\% & 51.2\% & 11.6\% \\
                Dynamic & 15090 & 18530 & \textbf{26019} & 34.9\% & 12.6\% \\
                Age & 11075 & 14756 & 11076 & \textbf{16811} & 9.8\% \\
                Kinematic & 3008 & 3360 & 3536 & 2002 & \textbf{5663} \\
                \hline
        \end{tabular}
        \label{tab:thick pairs}
        \tablefoot{Numbers on the diagonal are the total number of stars selected for that method, marked in bold. The upper half contains the fraction of the smaller of the two totals, giving a comparison to the maximum overlap. The kinematic method is applied to a smaller space and therefore includes fewer stars.}
\end{table}

\begin{figure*}
\centering
\includegraphics[width=\hsize]{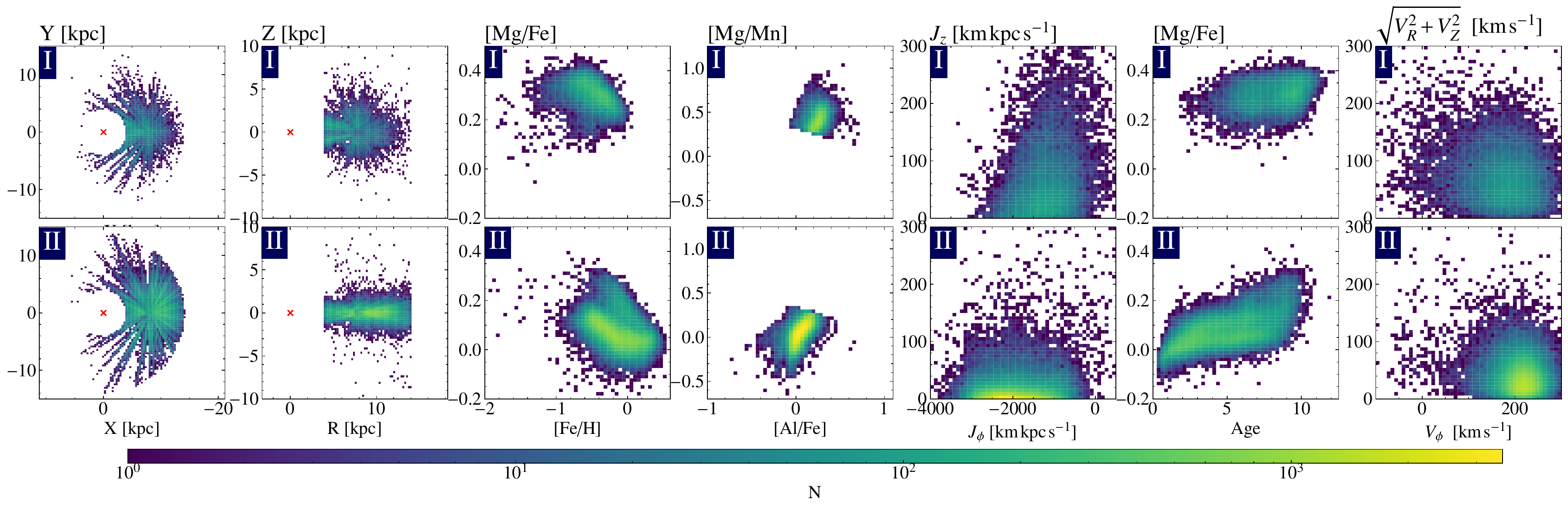}
\includegraphics[width=\hsize]{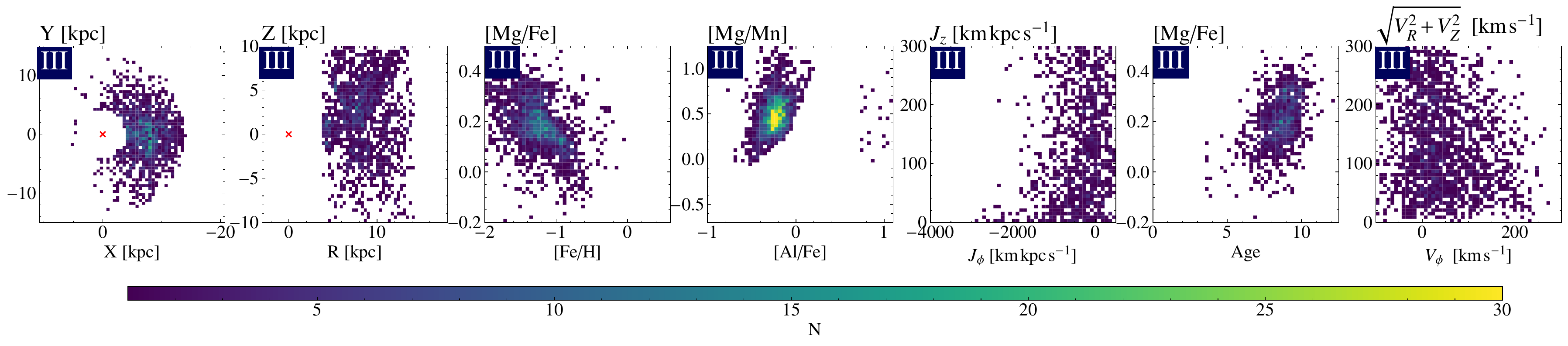}
  \caption{Distribution of stars from the [Mg/Mn]-[Al/Fe] selection in Sect. \ref{section:chem_haywood}.
  \textit{Left to right}: X-Y location, $ R_{\mathrm{gal}} $-Z location, [Mg/Fe]-[Fe/H], [Mg/Mn]-[Al/Fe], $J_\phi$-$J_z$, and Age-[Mg/Fe].
  \textit{Top}: Thick disc. \textit{Middle}: Thin disc. \textit{Bottom}: Halo.
          }
     \label{fig:mgmnalfe results full}
\end{figure*}

Figure \ref{fig:mgmnalfe results full} shows the results from the chemical selection using the [Mg/Mn]-[Al/Fe] plane in Sect. \ref{section:chem_haywood}, in which we can see stars from Region III occupying the low-metallicity region, with [Fe/H] $ < -0.5 $ except for some scatter.
Stars in Region II, the thin disc, are mostly confined to within approximately 2 kpc of the mid-plane, while stars from Region I, the thick disc, are found approximately 5 kpc of the mid-plane, and stars from Region III are spread further than 10 kpc.
Region I contains no stars younger than 2 Gyr, while Region II contains almost no stars older than 10.5 Gyr.
The majority of stars in Region II are in the area with $J_z <$ 100 km kpc s$^{-1}$.

\begin{figure*}
\centering
\includegraphics[width=\hsize]{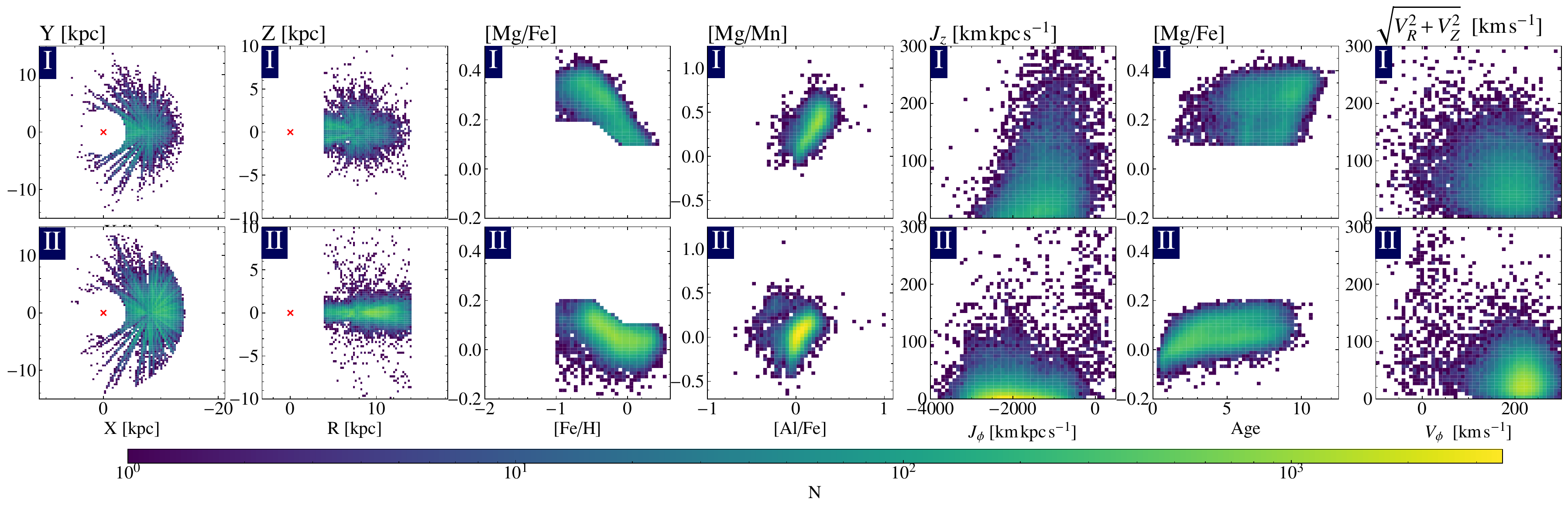}
\includegraphics[width=\hsize]{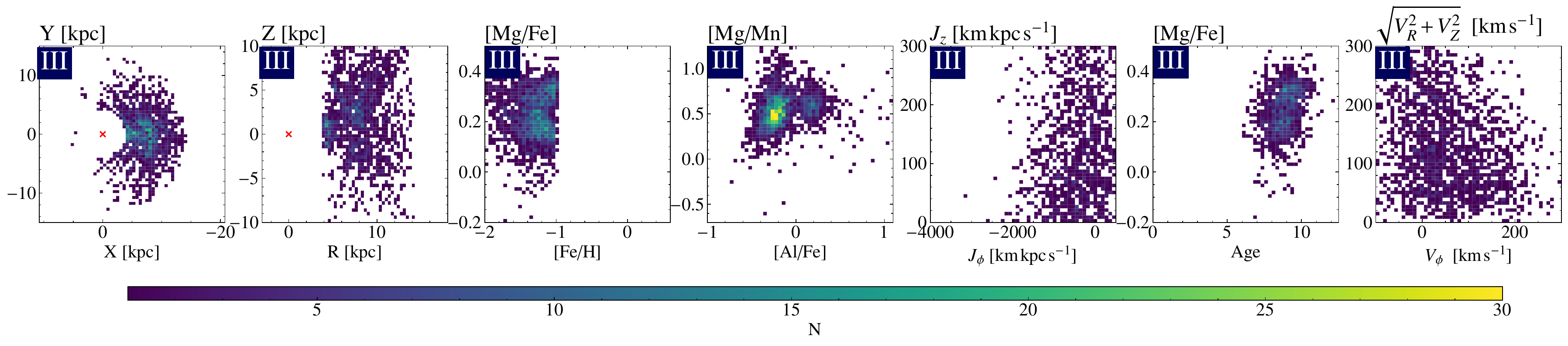}
  \caption{Same as Fig. \ref{fig:mgmnalfe results full} but for the [Mg/Mn]-[Al/Fe] selection (Sect. \ref{section:chem_simple}).
          }
     \label{fig:simple results full}
\end{figure*}

Figure \ref{fig:simple results full} shows the results from the chemical selection using the [Mg/Fe]-[Fe/H] plane in Sect. \ref{section:chem_simple}.
We see stars from Region III concentrated into two regions of the [Mg/Mn]-[Al/Fe] plane.
The biggest concentration is centred at approximately ([Mg/Mn] $ = 0.5$, [Al/Fe] $ = -0.2$), and a smaller one is centred at approximately ([Mg/Mn] $ = 0.6$, [Al/Fe] $ = 0.2$).
Stars in Region II, the thin disc, are mostly confined to within about 2.5 kpc of the mid-plane with some scatter, while stars from Region I, the thick disc, are found within about 5 kpc of the mid-plane, and stars from Region III are spread evenly and further than 10 kpc from the mid-plane.
Stars in Region II The $J_\phi$-$J_z$ plane at $ -3000 $ km kpc s$^{-1} < J_\phi < -1500 $ km kpc s$^{-1}$ seem to extend to larger values of $J_z$, approximately 120 km kpc s$^{-1}$ instead of approximately 80 km kpc s$^{-1}$.
The cut in [Mg/Fe] is visible in the Age-[Mg/Fe] plane, where we also see very few stars younger than 2 Gyr in Region I, and very few stars older than 9.5 Gyr in Region II.
Region III contains no stars younger than 6 Gyr.

\begin{figure*}
\centering
\includegraphics[width=\hsize]{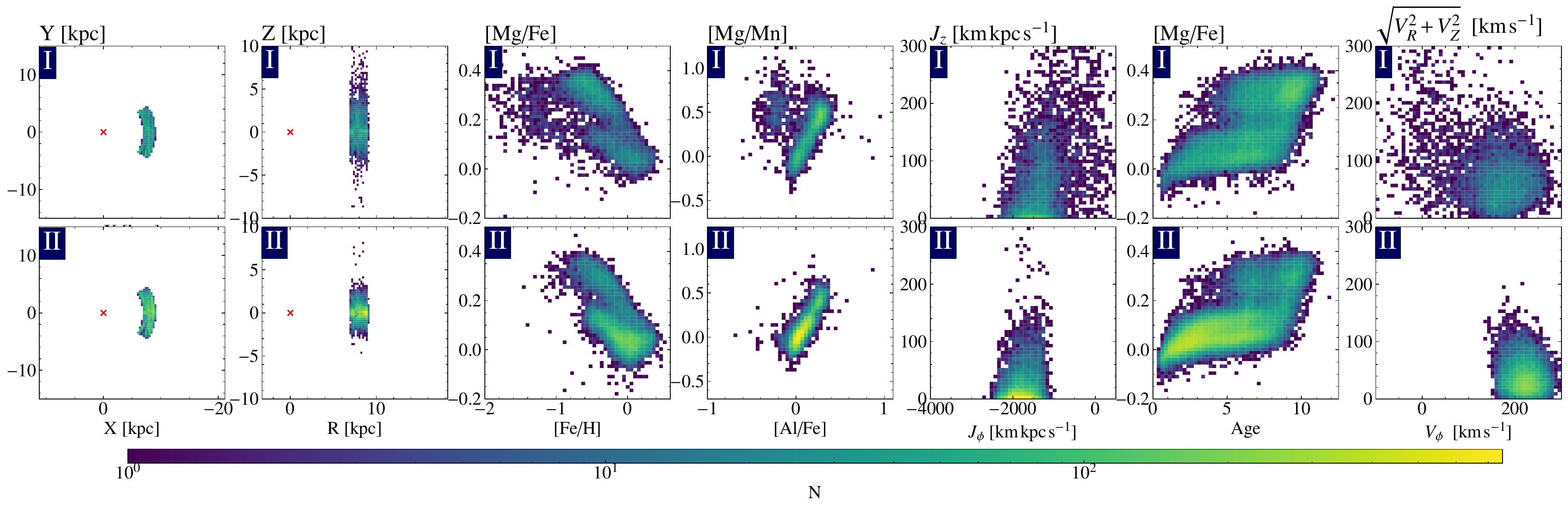}
\includegraphics[width=\hsize]{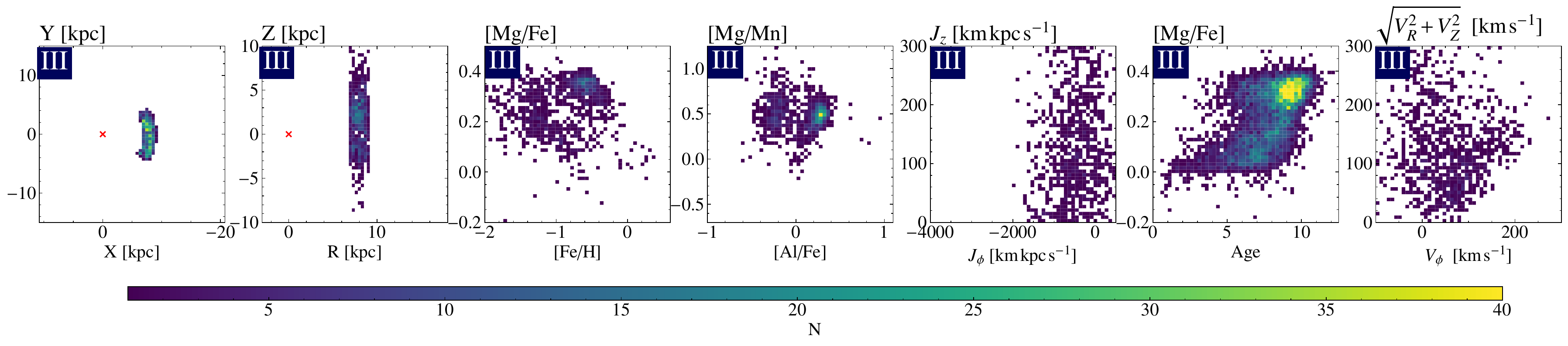}
  \caption{Distribution of stars from the kinematic selection in Sect. \ref{section:kinematic_method}.
  \textit{Left to right}:  X-Y location, $ R_{\mathrm{gal}} $-Z location, [Mg/Fe]-[Fe/H], [Mg/Mn]-[Al/Fe], $J_\phi$-$J_z$,  Age-[Mg/Fe], and Toomre plots.
  \textit{Top}: Thick disc. \textit{Middle}: Thin disc. \textit{Bottom}: Halo
          }
     \label{fig:kinematic results full}
\end{figure*}

Figure \ref{fig:kinematic results full} shows the results from the kinematic selection in Sect. \ref{section:kinematic_method}.
Stars in Region II, the thin disc, are mostly confined to within approximately 2.5 kpc of the mid-plane, while stars from Region I, the thick disc, are dense out to about 4 kpc but found in lower numbers out to 10 kpc from the mid-plane, and stars from Region III are spread out to 10 kpc but with lower density throughout.
Region III contains a small clump of stars from the high-$ \alpha $ sequence, centred at approximately ([Mg/Fe] $ = 0.35$, [Fe/H] $ = -0.6$) and ([Mg/Mn] $ = 0.5$, [Al/Fe] $ = -0.15$).
Stars in Region I seem to be mostly contained within $ -2300 $ km kpc s$^{-1} < J_\phi < -600 $ km kpc s$^{-1}$, with some scatter at larger $ J_\phi $.
Stars in Region II are more tightly confined within $ -2500 $ km kpc s$^{-1} < J_\phi < -1000 $ km kpc s$^{-1}$, with no scatter.
Stars in Region III are scattered across $ J_\phi > -2000 $ km kpc s$^{-1}$.
Regions I and II have stars in the same shape in the Age-[Mg/Fe] plane, but Region I have the densest part at Age = 9.5 Gyr and [Mg/Fe] = 0.3, while Region II has the entire area with [Mg/Fe] $< 0.18 $ be higher density.
Region III also has the densest concentration of stars at Age = 9.5 Gyr and [Mg/Fe] = 0.3.

\begin{figure*}
\centering
\includegraphics[width=\hsize]{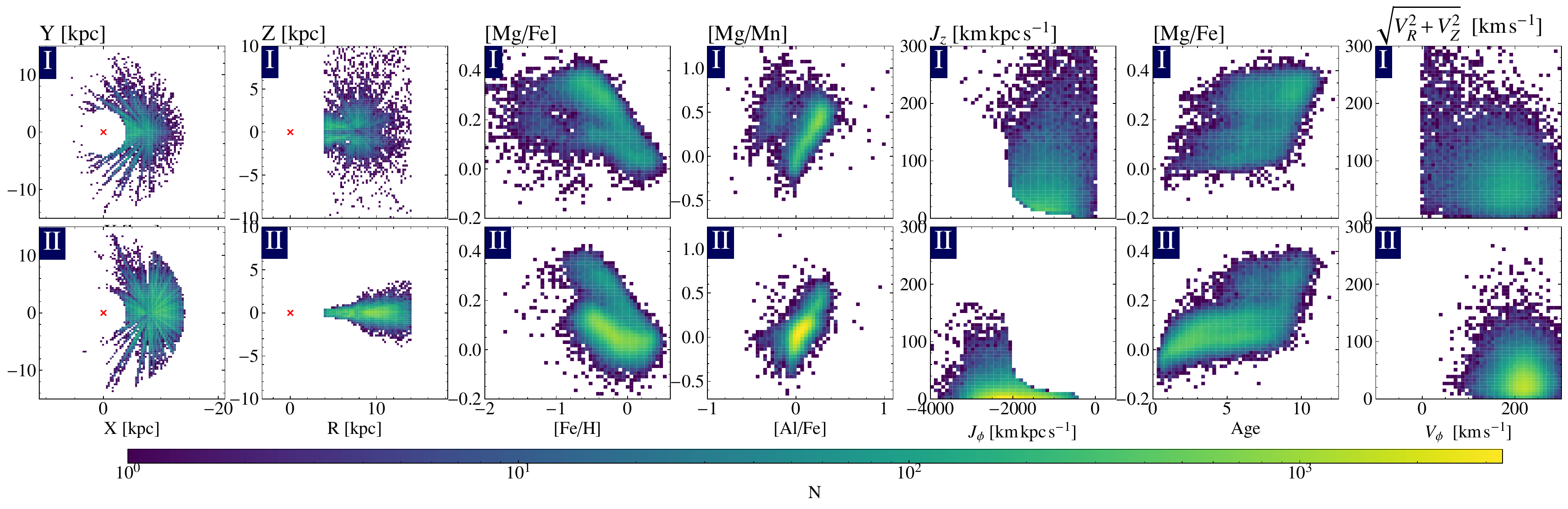}
\includegraphics[width=\hsize]{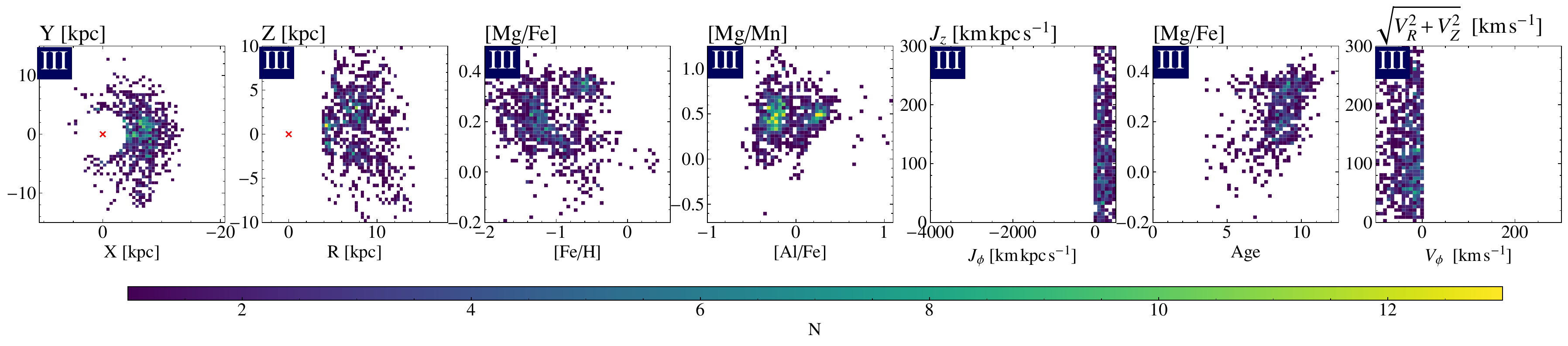}
  \caption{Same as Fig. \ref{fig:mgmnalfe results full} but for the dynamic selection (Sect. \ref{section:dynamic_method}).
          }
     \label{fig:dynamic results full}
\end{figure*}

Figure \ref{fig:dynamic results full} shows the results from the dynamical selection in Sect. \ref{section:dynamic_method}.
We see that Region I, the thick disc, has a clear underdensity of stars running along the plane of the Galaxy for the length of the disc.
Its highest density areas are two small triangular regions about 1 kpc from the place between the inner edge at $R_{\mathrm{gal}} = 4$ kpc and $R_{\mathrm{gal}} = 8$ kpc, and the density noticeably decreases beyond $ R_{\mathrm{gal}} = 10 $ kpc.
We also see that in the $ R_{\mathrm{gal}} $-Z plot, Region II, the thin disc, is mostly confined to a wedge shape of less than 1 kpc in height at the inner edge at $R_{\mathrm{gal}} = 4$ kpc and extending to 4 kpc from the mid-plane at the outer edge at $R_{\mathrm{gal}} = 16$ kpc.
Region III occupies the regions categorised as accreted ([Fe/H] $ > -1 $) and a small but distinct group of high-$ \alpha $ sequence stars located at approximately ([Mg/Fe] $ = 0.35$, [Fe/H] $ = -0.6$) and ([Mg/Mn] $ = 0.5$, [Al/Fe] $ =0.2$).
Stars in both Region I and II are found across the full range of ages, but Region I has the highest density at about $ 9.5 $ Gyr, while Region II show high density from $ 0 $ to about $ 7.5 $ Gyr.
Due to the selection criteria for Region III, all stars have $ V_\phi > 0 $ km s$^{-1}$ by construction, and appear uniformly distributed in the $ \sqrt{V^2_R + V^2_Z} $ direction.

\begin{figure*}
\centering
\includegraphics[width=\hsize]{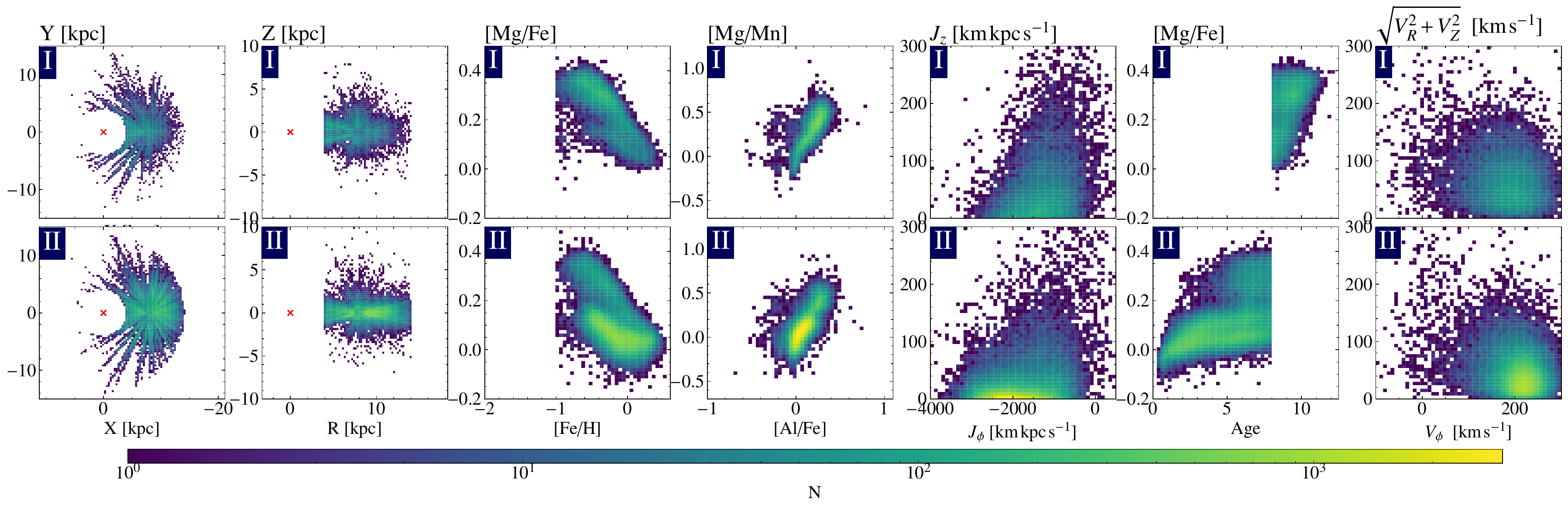}
\includegraphics[width=\hsize]{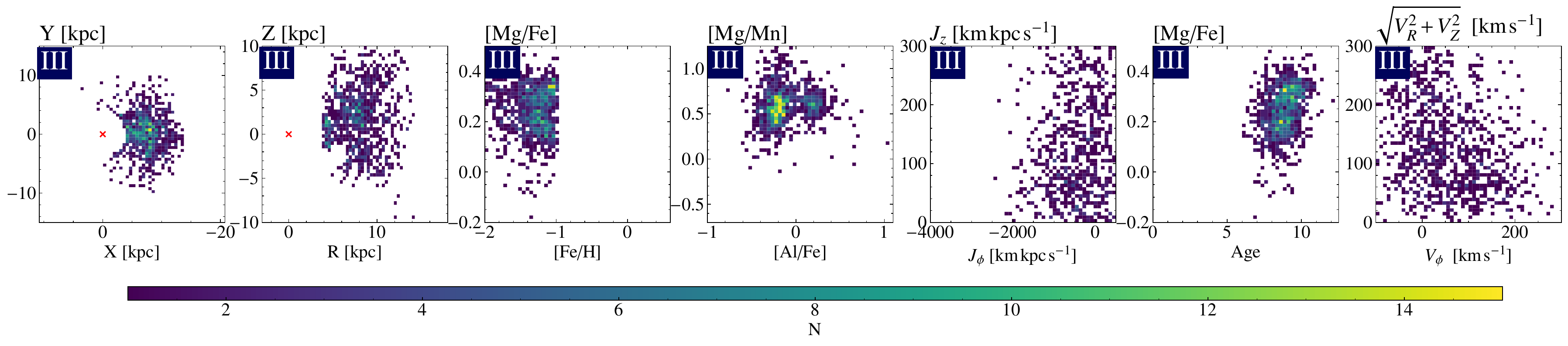}
  \caption{Same as Fig. \ref{fig:mgmnalfe results full} but for the age-based selection (Sect. \ref{section:age_method}).}
     \label{fig:age results full}
\end{figure*}

Figure \ref{fig:age results full} shows the results from the age-based selection in Sect. \ref{section:age_method}.
We see that Region III includes, apart from the accreted stars with low Al, a small part of the thick disc in the [Mg/Mn]-[Al/Fe] plane with a group of stars centred at approximately ([Mg/Mn] $ = 0.6$, [Al/Fe] $ = 0.2$).
We also see that stars in Region II, the thin disc, are mostly confined to within about 2 kpc of the mid-plane, while stars from Region I, the thick disc, are found approximately within 4 kpc of the mid-plane and decrease in density with radial distance, and stars from Region III are spread evenly and are present out to 10 kpc.
Stars in Region II reach $ J_z $ of about 200 km kpc s$^{-1}$, with some scatter at higher values, even at $J_\phi > -2000 $ km kpc s$^{-1}$, which is an area associated with the thick disc.

\begin{figure*}
        \centering
        \includegraphics[width=\hsize]{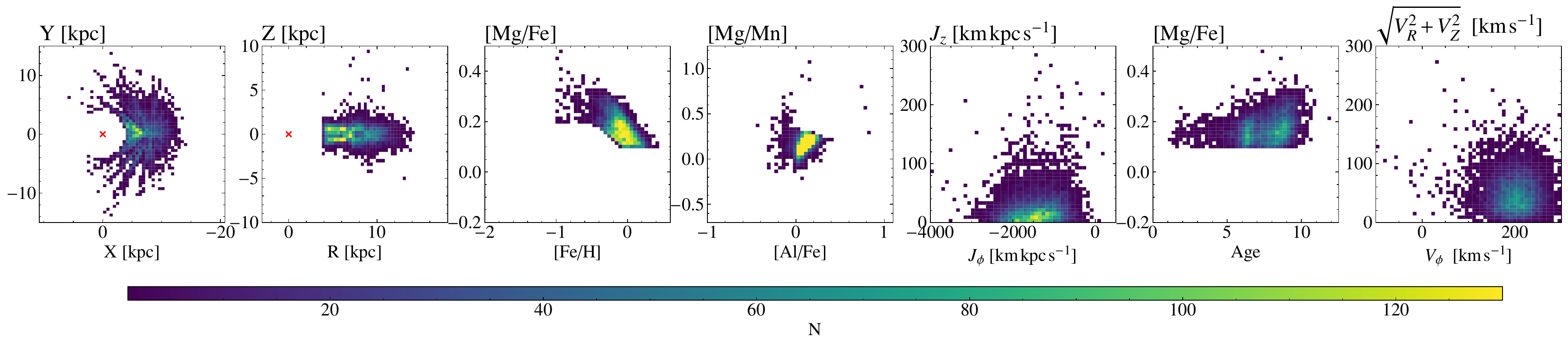}
        \caption{Distribution of stars selected as thick disc (high-alpha) by the [Mg/Fe]-[Fe/H] method but not by the [Mg/Mn]-[Al/Fe] method.
                \textit{Left to right}:  X-Y location, $ R_{\mathrm{gal}} $-Z location, [Mg/Fe]-[Fe/H], [Mg/Mn]-[Al/Fe], $J_\phi$-$J_z$, and Age-[Mg/Fe].
        }
        \label{fig:special0}
\end{figure*}

\begin{figure*}
        \centering
        \includegraphics[width=\hsize]{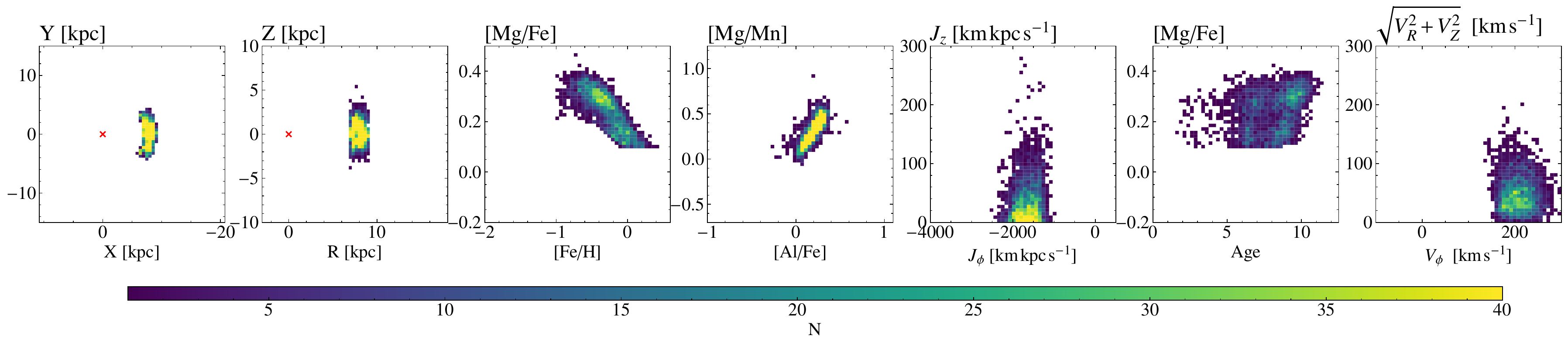}
        \caption{Distribution of stars selected as belonging to the thick disc (high-alpha) by the [Mg/Fe]-[Fe/H] method but to the thin disc by the kinematic method.
                \textit{Left to right}:  X-Y location, $ R_{\mathrm{gal}} $-Z location, [Mg/Fe]-[Fe/H], [Mg/Mn]-[Al/Fe], $J_\phi$-$J_z$, and Age-[Mg/Fe].
        }
        \label{fig:opposites}
\end{figure*}

Finally, Fig. \ref{fig:special0} shows the stars selected as thick disc (high-alpha) by the [Mg/Fe]-[Fe/H] method but not by the [Mg/Mn]-[Al/Fe] method, showing the differences between the two chemical methods.
These stars show mostly thin disc characteristics in their location, dynamics, and kinematics by being confined within 2 kpc of the Galactic midplane, having $ J_z > 100 $ kpc km s$^{-1}$, and forming a relatively tight group around $V_\phi = 200 $ km s$^{-1}$ in the Toomre plot.
Figure \ref{fig:opposites} shows the stars selected as thick disc (high-alpha) by the [Mg/Fe]-[Fe/H] method but thin disc by the kinematic method.

\end{appendix}

\end{document}